\documentclass[conference]{IEEEtran}
\IEEEoverridecommandlockouts
\usepackage{graphicx}%
\usepackage{adjustbox}%
\usepackage{array}
\usepackage{lipsum}
\usepackage{multirow}%
\usepackage{amsmath,amssymb,amsfonts}%
\usepackage{amsthm}%
\usepackage{enumitem}
\usepackage{mathrsfs}%
\usepackage{xcolor}%
\usepackage{textcomp}%
\usepackage{manyfoot}%
\usepackage{booktabs}%
\usepackage{algorithm}%
\usepackage{algorithmicx}%
\usepackage{algpseudocode}%
\usepackage{listings}%
\usepackage{tcolorbox}%
\usepackage{soul}
\usepackage[T1]{fontenc}
\usepackage{fix-cm}

\usepackage{graphicx}
\usepackage{float}
\usepackage{stfloats}
\usepackage{lipsum}

\usepackage{cite}%
\usepackage{array}%
\usepackage{booktabs}%
\usepackage{hyperref}%
\usepackage{caption}%
\usepackage{tabularx}%
\usepackage{float}

\usepackage{graphicx}
\usepackage[english]{babel}
\usepackage{float}
\usepackage{mdframed} 

\usepackage{longtable}%
\usepackage{multirow}%
\usepackage{rotating}%
\usepackage{graphicx}
\def\BibTeX{{\rm B\kern-.05em{\sc i\kern-.025em b}\kern-.08em
    T\kern-.1667em\lower.7ex\hbox{E}\kern-.125emX}}
\begin{document}

\title{Microservice Vulnerability Analysis: A Literature Review with Empirical Insights}

\author{

\IEEEauthorblockN{Raveen Kanishka Jayalath\textsuperscript{*}}
\IEEEauthorblockA{\textit{University of Adelaide, Australia}\\
raveenkanishka.jayalath@student.adelaide.edu.au}

\and
\IEEEauthorblockN{Hussain Ahmad\textsuperscript{* $\dagger$} \thanks{*Authors contributed equally to this work.} \thanks{\textsuperscript{ $\dagger$}Corresponding author.}}
\IEEEauthorblockA{\textit{University of Adelaide, Australia}\\
hussain.ahmad@adelaide.edu.au}
\and
\IEEEauthorblockN{Diksha Goel}
\IEEEauthorblockA{\textit{CSIRO's Data61, Australia}\\
diksha.goel@data61.csiro.au}
\and
\IEEEauthorblockN{\hspace{3cm}Muhammad Shuja Syed}
\IEEEauthorblockA{\hspace{3cm}\textit{SLB, USA} \\
\hspace{3cm}Msyed13@slb.com}
\and
\IEEEauthorblockN{Faheem Ullah}
\IEEEauthorblockA{\textit{University of Adelaide, Australia}\\
faheem.ullah@adelaide.edu.au}

}

\maketitle

\pagestyle{plain}

\begin{abstract}

Microservice architectures are revolutionizing both small businesses and large corporations, igniting a new era of innovation with their exceptional advantages in maintainability, reusability, and scalability. However, these benefits come with significant security challenges, as the increased complexity of service interactions, expanded attack surfaces, and intricate dependency management introduce a new array of cybersecurity vulnerabilities. While security concerns are mounting, there is a lack of comprehensive research that integrates a review of existing knowledge with empirical analysis of microservice vulnerabilities. This study aims to fill this gap by gathering, analyzing, and synthesizing existing literature on security vulnerabilities associated with microservice architectures. Through a thorough examination of 62 studies, we identify, analyze, and report 126 security vulnerabilities inherent in microservice architectures. This comprehensive analysis enables us to (i) propose a taxonomy that categorizes microservice vulnerabilities based on the distinctive features of microservice architectures; (ii) conduct an empirical analysis by performing vulnerability scans on four diverse microservice benchmark applications using three different scanning tools to validate our taxonomy; and (iii) map our taxonomy vulnerabilities with empirically identified vulnerabilities, providing an in-depth vulnerability analysis at microservice, application, and scanning tool levels. Our study offers crucial guidelines for practitioners and researchers to advance both the state-of-the-practice and the state-of-the-art in securing microservice architectures.

\end{abstract}

\begin{IEEEkeywords}
microservice, cybersecurity, taxonomy, vulnerability scanning, vulnerability analysis
\end{IEEEkeywords}

\section{Introduction}\label{sec1}

Microservice architectures have become increasingly popular as both small and large enterprises shift from monolithic structures to microservices to enhance service quality \cite{blinowski2022monolithic, luo2024optimizing}. Unlike traditional monolithic architectures, microservice architectures decompose complex applications into independent, loosely coupled services known as microservices \cite{pimentel2021self}. Each microservice performs a distinct function and communicates with other services via lightweight interfaces like HTTP resource APIs \cite{velepucha2023survey}. This method facilitates faster application delivery and boosts reliability, as each microservice can be developed, deployed, and scaled independently \cite{gorden2023predicting}. Microservices are generally deployed within software containers, and managing these containers during runtime often involves container orchestration platforms such as Red Hat OpenShift, Kubernetes, and Docker Swarm \cite{ahmad2024smart}.

While microservices offer numerous benefits, they also introduce distinct security vulnerabilities due to their fundamental characteristics, such as service containerization, inter-service communication protocols, and container orchestration platforms \cite{abdulsatar2024towards, mateus2021security}. Additionally, the inclusion of third-party components within microservice architectures brings about new cybersecurity risks, further expanding the attack surface available to malicious actors \cite{dragoni2017microservices}. The landscape of security vulnerabilities in microservice architectures is constantly evolving, posing significant challenges for organizations \cite{shmeleva2020microservices}. As companies increasingly adopt microservice architectures, they face a growing array of sophisticated threats that can exploit weaknesses within their systems. These vulnerabilities can have severe impacts, including data breaches, service disruptions, and financial losses \cite{mateus2021security}. In particular, the decentralized nature of microservices, while offering benefits in scalability and flexibility, also introduces new attack surfaces \cite{nkomo2019software}. Each microservice can potentially become a point of entry for malicious actors, leading to cascading effects across the entire system. The interconnectivity of microservices means that a compromise in one service can quickly spread, affecting the integrity and availability of other services \cite{dias2020microservices}. Beyond financial repercussions, cyberattacks exploiting vulnerabilities in microservice architectures can undermine customer trust, tarnish reputations, and lead to regulatory penalties \cite{yu2019survey}.

Given the continuously evolving landscape of security vulnerabilities in microservice architectures \cite{shmeleva2020microservices}, developing a comprehensive security vulnerability taxonomy is essential. Such a taxonomy identifies and classifies the diverse and complex vulnerabilities unique to microservice architectures, enhancing understanding and communication among security professionals, developers, and stakeholders. It enables more accurate risk assessments and targeted mitigation strategies, ensuring efficient resource allocation and robust security measures. Additionally, it facilitates compliance with regulatory requirements and provides a benchmark for continuous improvement. Therefore, in this study, we aim to develop a taxonomy of security vulnerabilities associated with microservice architectures. To achieve this, we collect, analyze, and synthesize findings from 62 relevant studies from the existing literature that report on the security vulnerabilities of microservice architectures. From these studies, we extract 126 distinct vulnerabilities. To ensure the taxonomy's effectiveness, we categorize these vulnerabilities based on the unique characteristics of microservices, such as those related to API gateways, service discovery mechanisms, and containerization. By focusing on these specific aspects, our taxonomy not only highlights the critical areas where microservice architectures are most vulnerable but also provides a detailed understanding of addressing these risks. This categorization helps in pinpointing precise security measures required for different components of the architecture. For instance, vulnerabilities related to API gateways might necessitate stringent access controls and monitoring, while those associated with containerization could require robust isolation and orchestration practices.

In addition to developing the taxonomy, we also validate it through empirical analysis using four benchmark microservice applications. To ensure the robustness and accuracy of our taxonomy, we employ three different vulnerability scanning tools to detect vulnerabilities within these applications. The detected vulnerabilities are then meticulously mapped to the categories within our taxonomy. This rigorous process not only confirms the relevance and comprehensiveness of our taxonomy but also provides several significant insights. Firstly, it underscores the criticality of specific vulnerabilities, enabling us to identify which types pose the greatest risks to microservice architectures. These insights are invaluable for both researchers and practitioners. For researchers, the findings highlight potential areas for future study, suggesting new avenues for improving security measures and developing more advanced detection tools. For practitioners, the insights offer practical guidance on prioritizing their security efforts, focusing on the most critical vulnerabilities to enhance the overall security posture of their microservice environments. By validating our taxonomy through empirical analysis and mapping detected vulnerabilities, we not only confirm its accuracy but also enrich the understanding of the security landscape in microservice architectures. This dual approach ensures that our taxonomy is not only theoretically sound but also practically applicable, providing a robust framework for addressing the complex and evolving security challenges in microservice environments. In summary, our study makes the following contributions:

\begin{itemize} [leftmargin=*]
    \item We propose a comprehensive taxonomy that identifies and classifies the diverse and complex security vulnerabilities for microservice architectures. This taxonomy enhances the understanding of microservice vulnerabilities among security professionals, developers, and stakeholders, providing a structured framework for more accurate risk assessments and targeted mitigation strategies.
    \item we conduct an empirical analysis on four benchmark microservice applications. Using three different vulnerability scanning tools, we detect a range of vulnerabilities within these applications. This rigorous analysis not only confirms the relevance and comprehensiveness of our taxonomy but also provides real-world insights into the types and severity of microservice vulnerabilities.
    \item We map the detected vulnerabilities from our empirical analysis to those within our taxonomy. This mapping uncovers several insights into the criticality of microservice vulnerabilities and provides essential guidelines for practitioners and researchers to advance both practical and theoretical aspects of securing microservice architectures.
\end{itemize}

The rest of the paper is structured as follows: Section \ref{sec2} discusses the background and related work concerning the security vulnerabilities of microservice architectures. Section \ref{sec3} outlines the research methodology used to extract 62 relevant studies from existing literature on microservice vulnerabilities. Section \ref{sec4} presents the proposed taxonomy, while Section \ref{sec5} details the empirical analysis. Section \ref{sec6} explains the mapping process of the detected vulnerabilities to the taxonomy, discusses the insights gained, and provides guidelines for researchers and practitioners. Section \ref{sec7} addresses threats to validity, and Section \ref{sec8} concludes the paper. Additionally, the repository of this study is available at \cite{repository2024} for further reference, including access to data, supplementary materials, and implementation details.

\section{Background and Related Work}\label{sec2}

In this section, we describe monolithic and microservice architectures and compare them, as well as discuss the cybersecurity landscape for both small and large-scale enterprises.

\subsection{Microservice Architecture}

Microservices are a contemporary architectural style designed to structure complex systems as a collection of small, loosely coupled software components known as microservices \cite{hannousse2021securing}. Each microservice typically handles a specific function or a segment of the business logic of an application \cite{de2019monolithic}. This architecture allows for using different programming languages, varied database technologies, and isolated testing of individual components, minimizing impact on the overall system \cite{salah2016evolution}. Microservices can communicate either directly through HTTP APIs or indirectly via message brokers. They can be deployed using virtual machines or, more commonly, lightweight containers, which are preferred for their ease of use, cost-effectiveness, and rapid startup and execution times \cite{waseem2020systematic}.

The adoption of microservices improves software quality by enhancing maintainability, interoperability, reusability, and scalability \cite{ahmad2024smart}. With the right distributed platforms and technologies, microservices can be deployed, scaled, replaced, and removed independently without disrupting system availability \cite{gorden2023predicting, goel2023evolving}. Each microservice is designed to handle a specific business capability, making it reusable across various applications and domains \cite{surianarayanan2019essentials}. Microservices differ from monolithic architectures and traditional service-oriented architectures primarily in their smaller size, scalability, and autonomy of each service \cite{de2019monolithic}. Therefore, the adoption of microservices is becoming increasingly widespread across industries of all sizes, from Small and Medium-sized Enterprises (SMEs) businesses to large enterprises, due to their unmatched characteristics \cite{andriyanto2019adopting, sani2019development, hannousse2021securing}. Major organizations such as Coca-Cola, Netflix, eBay, and Amazon are at the forefront of this trend \cite{luo2024optimizing}.

\subsection{Security in Microservice Architectures}

Microservices, while offering numerous benefits, also introduce significant security challenges compared to traditional monolithic applications \cite{shmeleva2020microservices}. The decentralization of functionality into multiple microservices expands the attack surface, as each individual service can potentially be a target for security breaches \cite{dias2020microservices, kalubowila2021optimization}. This expansion makes comprehensive application security more difficult because securing a microservice architecture requires a holistic approach, rather than relying on isolated defenses for each service \cite{torkura2018cyber}. Unlike monolithic systems where security measures can be applied uniformly, microservices necessitate coordinated security strategies across the entire ecosystem \cite{hannousse2021securing}. In a microservice architecture, the network-based communication between services introduces a potential attack surface. If a microservice is compromised, it could issue harmful requests to other services within the network \cite{yu2019survey}. To address this, TLS is often used to encrypt the communications between microservices \cite{yu2019survey}. Additionally, platforms like Kubernetes provide mechanisms for inter-service authorization to regulate valid access permissions \cite{dell2023kubehound}. However, challenges remain in synchronizing these authorization rules across the distributed environment and updating them when services or users are added or removed \cite{de2022authentication}. Moreover, inter-service authorization must align with conceptual models of architectures and follow the principle of least privilege \cite{he2017authentication}. 

Authentication in microservice architectures introduces similar challenges, particularly concerning the storage of authentication information \cite{de2022authentication}. If this information is centralized in an authentication server, the server needs to be updated whenever new microservices or users are added \cite{shulin2020research}. Conversely, if each microservice retains its own authentication data, then each service must be updated individually to reflect any changes \cite{shulin2020research}. The process of decomposing an application into microservices adds complexity, impacting both scalability and security \cite{he2017authentication}. Moreover, microservice architectures can be vulnerable due to their reliance on images from public or private repositories, which may contain potential infections. Besides exploiting application vulnerabilities, attackers can also target misconfigurations within microservices \cite{habbal2020enhancing}. A breach involving compromised credentials, services, networks, or hosts can potentially lead to the complete compromise of the entire cluster of microservices \cite{hannousse2021securing}.

\textbf{Security State of SMEs.} Small and medium-sized enterprises adopting microservice architectures often face significant security challenges due to the complexity of microservices and their limited resources. The expanded attack surface and intricate inter-service communications inherent in microservices require robust security measures \cite{shmeleva2020microservices}, which can be difficult for SMEs with constrained budgets and expertise. To mitigate these risks, SMEs are focusing on implementing foundational security practices, utilizing cloud services for enhanced protection, and leveraging monitoring and container security solutions. For instance, Stone \& Chalk \cite{stoneandchalk}, a leading innovation hub in Australia, has partnered with AustCyber \cite{austcyber} to bolster the cybersecurity capabilities of SMEs within the hub \cite{merger}. A notable example is Migrova \cite{Migrova}, an innovative startup within Stone \& Chalk, which provides an immigration marketplace connecting migrants with Australian migration agents. Migrova is developing its software architecture with a security-by-design approach to protect the sensitive information of both migrants and migration agents \cite{Migrova}. These efforts highlight the pressing need for comprehensive vulnerability analysis of microservice architectures to effectively address the evolving threats faced by SMEs in this dynamic landscape.

\textbf{Security State of Large Enterprises.} The architectural transformation has elevated cybersecurity to a matter of national and global security, requiring a comprehensive and coordinated approach to mitigate potential threats to economic stability, national sovereignty, and global supply chains \cite{tvaronavivciene2020cyber} \cite{uddin2020cybersecurity} \cite{ahmad2023review}. Notable cyber attacks, such as those on Saudi Aramco (2012), Norwegian oil companies (2014), Colonial Pipeline (2021), and Northern European Refineries (2022) \cite{olech2021cybersecurity} \cite{jacobs2016industrial} \cite{beerman2023review} \cite{stergiopoulos2020cyber}, demonstrate the critical importance of cybersecurity of energy sectors. The Bangladesh Bank heist highlights the vulnerability of an entire country's economy to cyber threats \cite{mazumder2020spillover}. Additionally, attacks on Mondelez International (2021), Toyota (2022), and Colorox (2023) underscore the vulnerability of manufacturing to cybersecurity risks \cite{tatar2021digital} \cite{rahman2023manufacturing} \cite{elsaleiby2024moderation}. As software development increasingly adopts microservice architectures, investigating the security vulnerabilities of such architectures becomes increasingly vital. Software developers must recognize that any oversight or misconception leads to damages that extend beyond individuals and organizations, emphasizing the need for a comprehensive and coordinated approach to microservice security \cite{jarmon2018cyber}.

\subsection{Related Work}

In the existing research on the security of microservice architectures, studies typically fall into two main categories: literature reviews and empirical analysis. Existing literature reviews provide valuable overviews of vulnerabilities associated with microservices, cataloging known issues and summarizing previous findings \cite{haindl2024systematic, berardi2022microservice, de2022authentication, hannousse2021securing, ponce2022smells, pereira2021security}. These reviews often offer a broad perspective on the security challenges inherent in microservice environments. Empirical analysis, on the other hand, focus on the practical identification and evaluation of vulnerabilities within microservice architectures \cite{torkura2017integrating, chondamrongkul2020automated, abdulsatar2024towards, tai2023framework, baker2019novel}. These studies often propose frameworks for vulnerability assessment and management specifically designed for microservices, aiming to structure and address security concerns through methodical approaches.

Despite the wealth of literature available, there is a notable gap in research that combines both comprehensive literature reviews and empirical analysis within a single study. Most existing studies focus on one aspect without integrating both theoretical and practical perspectives. Our research addresses this gap by not only providing an extensive literature review but also developing a novel taxonomy of microservice vulnerabilities. This taxonomy is then empirically validated through practical analysis of microservice benchmark applications. By combining these approaches, our study offers a more holistic understanding of microservice vulnerabilities, incorporating both theoretical frameworks and practical validation. This dual approach provides deeper insights into the nature of vulnerabilities and enhances the development of effective security measures, distinguishing our work from prior research efforts.

\section{Research Methodology}\label{sec3}

This section outlines the methodology used to systematize existing knowledge on security vulnerabilities in microservice architectures. We employ a five-step methodology, which is described as follows:

\subsection{Research Question}

This study specifically concentrates on the vulnerabilities that are inherent to microservice architectures. By focusing solely on these vulnerabilities, the research aims to address a single, critical question: \textit{What are the vulnerabilities inherent to microservice architectures?} This focused inquiry seeks to identify and understand the specific security weaknesses that arise from the unique characteristics of microservices.

\begin{figure}[t]
\renewcommand{\arraystretch}{1.2}
        \centering
        \includegraphics[width=0.42\textwidth]{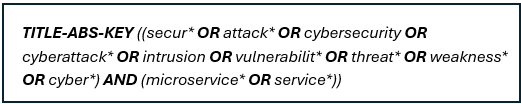}
        \caption{Search String.}
        \label{fig: search_string}
\end{figure}

\subsection{Search Strategy}

Our approach to finding and retrieving relevant studies on microservice vulnerabilities is executed in two phases. Initially, we create a search string to identify pertinent literature, as shown in Figure \ref{fig: search_string}. This search string is split into two components: the first component targets articles related to "vulnerabilities" and its synonyms, while the second component focuses on articles concerning "microservice architectures" and related terms. After conducting a pilot search using this string, we successfully retrieved articles and studies recognized for their relevance to vulnerabilities in microservice architectures. The search string, detailed in Figure \ref{fig: search_string}, effectively aligned with titles, keywords, and abstracts across all digital libraries through automated searching.

In the second phase, we explore seven digital libraries to maximize the number of relevant papers and articles, as shown in Table \ref{tab:publication_sources}. By applying the search string to these databases (excluding SpringerLink), we retrieve relevant papers based on titles, abstracts, and keywords. SpringerLink is excluded due to its limitation in searching specific sections such as titles, abstracts, and keywords within papers/articles \cite{braun2006using}. The SpringerLink engine typically applies search terms to either the titles of articles or their full texts. Searching by title identifies only a small number of relevant articles. However, searching the full text yields a much larger number of results, totaling 1,204 articles. To address this issue, we analyze the first 1000 papers from the total retrieved. This approach ensures that no relevant articles are excluded, as the Scopus library, which includes many articles indexed in SpringerLink, is also part of our search. Additionally, we do not use Google Scholar directly as a data source due to its low precision, which often results in a high volume of irrelevant papers.

\begin{table}[t]
    \centering
    \caption{Data Sources.}
    \label{tab:publication_sources}
    \begin{tabular}{p{3cm}p{4cm}}
        \hline
        \textbf{Source} & \textbf{URL} \\
        \hline
        IEEE Xplore & \url{https://ieeexplore.ieee.org} \\
        \hline
        ScienceDirect & \url{https://www.sciencedirect.com} \\
        \hline
        Scopus & \url{https://www.scopus.com} \\
        \hline
        Wiley & \url{https://onlinelibrary.wiley.com} \\
        \hline
        SpringerLink & \url{https://link.springer.com} \\
        \hline
        ACM & \url{https://dl.acm.org} \\
        \hline
         Web of Science & \url{https://www.webofknowledge.com} \\
        \hline
    \end{tabular}
\end{table}

\subsection{Inclusion and Exclusion Criteria}

We establish inclusion and exclusion criteria, as outlined in Table \ref{tab:criteria}, and apply these criteria to all extracted articles and papers. Only peer-reviewed papers and articles focused on vulnerabilities within microservice architectures are included in the selection process. Studies are selected regardless of their publication date. However, keynote addresses, panel discussions, editorials, reviews, and tutorial summaries are excluded. Additionally, papers published in languages other than English are not considered. When two papers on the same topic are published in different venues (such as a conference and a journal), the more mature paper is selected. Duplicate studies found across different libraries are excluded.

\begin{table}[t]
\renewcommand{\arraystretch}{1.2}
    \centering
    \caption{Inclusion and Exclusion Criteria.}
    \label{tab:criteria}
    \begin{tabular}{p{8.5cm}}
        \hline
        \textbf{Inclusion Criteria} \\
        \hline
        \textit{I1}: A study related to the security of microservice architecture \\
        \hline
        \textit{I2}: A study is selected irrespective of its publication date \\
        \hline
        \textbf{Exclusion Criteria} \\
        \hline
        \textit{E1}: A study that is written in a language other than English \\
        \hline
        \textit{E2}: A study that is not peer-reviewed \\
        \hline
        \textit{E3}: A study that is not accessible by digital libraries or Google search \\
        \hline
    \end{tabular}
\end{table}

\begin{figure}[t]
        \centering
        \includegraphics[width=0.5\textwidth]{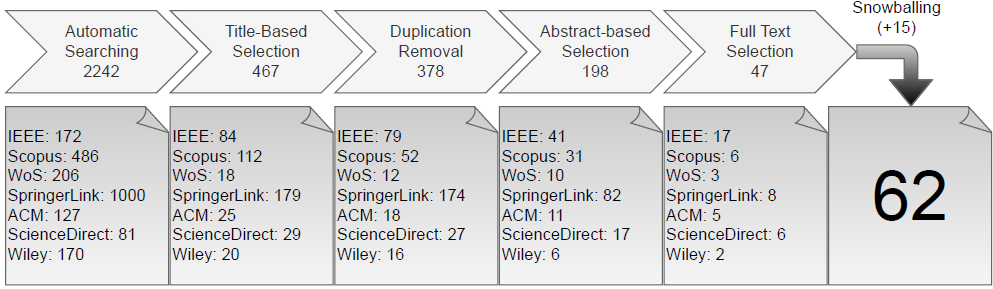}
        \caption{Study Selection Process.}
        \label{fig:Phases}
\end{figure}

\subsection{Study Selection}

We employ a six-step approach to select relevant studies from the existing literature. Figure \ref{fig:Phases} outlines this selection process, and we briefly describe each step as follows.

\begin{enumerate}[leftmargin=*]
    \item \textit{Automatic Search.} After applying the search string across eight digital libraries, the automated search retrieves 2242 potential papers and articles, including the first 1,000 papers from SpringerLink.
    \item \textit{Title-based Selection.} Following the automatic search, we filter the articles based on their titles. Articles with titles directly related to microservice vulnerabilities are included in the selection. Titles that are ambiguous or uncertain are carried forward to the next phase. This process results in a total of 467 articles remaining for further review.
    \item \textit{Duplication Removal.} Papers indexed in multiple libraries, such as Web of Science, Scopus, SpringerLink, ACM, IEEE Xplore, and ScienceDirect, are identified and removed. This process results in 378 unique studies remaining for review.
    \item \textit{Abstract-based selection.} We examine the abstracts and conclusions of the remaining articles to determine their relevance to vulnerabilities in microservice architectures. Applying the inclusion and exclusion criteria defined in Table \ref{tab:criteria} during this review reduces the number of relevant studies to 198.
    \item \textit{Full-text based selection.} We thoroughly read the full texts of the identified studies and applied the inclusion and exclusion criteria detailed in Table \ref{tab:criteria}. This process further refines the selection, resulting in 47 relevant studies.
    \item \textit{Snowballing.} We explore the references of the 47 identified studies and find 22 additional potential studies. We apply the inclusion and exclusion criteria to these articles, leaving 15 more papers for consideration. This brings the total number of relevant studies to 62.
    
\end{enumerate}

We then analyze the 62 studies to address our research question: \textit{What are the vulnerabilities inherent to microservice architectures?} In the following sections, we present the findings from this analysis, detailing the identified vulnerabilities and their implications for microservice security.

\section{Taxonomy of Microservice Vulnerabilities} \label{sec4}

In this section, we introduce a comprehensive taxonomy designed to categorize vulnerabilities in microservice architectures. Table \ref{tab:taxonomy} presents the proposed taxonomy of microservice vulnerabilities. Our taxonomy is structured into three tiers to provide a detailed and effective analysis of microservice vulnerabilities. The first tier covers broad domains of microservice architectures, such as DevOps practices, heterogeneous features, and microservice deployment and orchestration. The second tier delves deeper into specific features within these domains, including service discovery, identity management, and third-party dependencies. Finally, the third tier details the vulnerabilities associated with each specific feature of microservices. This structured approach provides researchers and practitioners with a clear framework for pinpointing vulnerabilities, enabling more targeted and effective security measures to enhance the resilience of microservice architectures against evolving threats.  

Through the analysis and synthesis of findings from the extracted 62 studies, we identified 126 vulnerabilities related to various features of microservice architectures. Table \ref{tab:taxonomy} categorizes these vulnerabilities by specific microservice features, organized into subcategories and the primary categories of our taxonomy. To enhance the real-world applicability of our taxonomy, we have mapped each identified vulnerability with its relevant Common Vulnerabilities and Exposures (CVE) identifier \cite{NVDdatabase}. CVE is a publicly available list that assigns standardized identifiers to known security vulnerabilities \cite{NVDdatabase}. Each CVE ID uniquely identifies a specific vulnerability, facilitating consistent tracking, reporting, and referencing across various platforms and tools. This mapping not only standardizes the identification process but also enables organizations and individuals to quickly assess the relevance and severity of vulnerabilities, prioritize mitigation efforts, and ensure robust security measures. By integrating CVE IDs into our taxonomy, we bridge the gap between theoretical analysis and practical application, making our categorization both comprehensive and actionable. In the following, we describe the primary taxonomy categories, their respective subcategories, and the specific vulnerabilities associated with them.

\subsection{\textbf{Network and Communication}}
Network and communication within microservice architectures refer to the essential infrastructure and protocols required to facilitate interactions and data exchanges across distributed microservices \cite{aksakalli2021deployment}. This is critically influential, impacting the security and operational efficiency of the entire microservice architecture. It includes different microservice features such as API Gateways and Endpoints, Service Discovery, and Network Segmentation and Isolation, each playing a vital role in the stability and security of microservices \cite{abdulsatar2024towards}. The importance of securing microservice communication networks cannot be overstated, as vulnerabilities lead to severe network compromises, unauthorized data access, and significant disruptions in operations, undermining the robustness of microservice architectures \cite{shmeleva2020microservices}. The Network and Communication class of vulnerabilities is divided into five subclasses, as discussed below:

\subsubsection{API Gateways and Endpoints}
API Gateways and Endpoints serve as critical components in microservice architectures, orchestrating the flow of data and managing service requests across distributed environments \cite{CarneiroSchmelmer2016Microservices}. These elements are crucial for enforcing security policies and managing network traffic, placing them at the heart of network security. Vulnerabilities within API gateways and endpoints, such as exposed API endpoints, can have far-reaching consequences, including extensive network compromise and unauthorized data access. Such breaches disrupt operational continuity and compromise the security of the entire network \cite{Garriga2018Taxonomy}. This subsection will detail the vulnerabilities unique to API Gateways and Endpoints, emphasizing their potential exploits and the resultant impacts on the network.

\newpage    

\onecolumn
{\fontsize{4.5pt}{5pt}\selectfont
\begin{longtable}{|m{5cm}|m{5cm}|m{4.5cm}|}
\caption{Taxonomy of Microservice Vulnerabilities.}\label{tab:taxonomy} \\
\hline
\centering \normalsize Primary Category & \centering  \normalsize Subcategory & \normalsize Microservice Vulnerability \\[2.5mm]
\hline
\endfirsthead
\hline
 \multirow{29}{*}{\small Network and Communication} & \multirow{6}{*}{\scriptsize API Gateways and Endpoints} & V$_{1}$: Exposed API Endpoints \cite{genfer2022avoiding, Garriga2018Taxonomy, RademacherSorgallaSachweh2018, genfer2021avoiding} \\
 &  & V$_{2}$: Accidental Exposure of Sensitive API Endpoints \cite{xu2019microservice, CarneiroSchmelmer2016Microservices}, \cite{prabath2020} \\
 &  & V$_{3}$: Untrusted Third-Party APIs \cite{chen2021systematic, Garriga2018Taxonomy, CarneiroSchmelmer2016Microservices, rudrabhatla2020security} \\
 &  & V$_{4}$: Weak API Authentication \cite{genfer2022avoiding, PereiraVale2021}, \cite{prabath2020}, \cite{Beyah2018SecurityPrivacy} \\
 &  & V$_{5}$: Insecure API Deserialization \cite{Alshuqayran2016Systematic, genfer2022avoiding}, \cite{Loureiro2021SecuringMicroservices, muresu2021investigating} \\
  &  & V$_{6}$: Misconfigured API Gateways \cite{xu2019microservice, DeAlmeida2022, RademacherSorgallaSachweh2018} \\
\cline{2-3}
  & \multirow{6}{*}{\scriptsize Service Discovery} & V$_{7}$: Service Registration Tampering \cite{BarrocaFilho2019MicroserviceHealth}, \cite{PriyaSethuramanKhan2023}, \cite{Zirpins2020ESOCC} \\
  &  & V$_{8}$: Unauthorized Service Discovery Access \cite{muresu2021investigating}, \cite{BarrocaFilho2019MicroserviceHealth}, \cite{PriyaSethuramanKhan2023} \\
 &  & V$_{9}$: Service Registration Validation Gaps \cite{PriyaSethuramanKhan2023} \\
 &  & V$_{10}$: Unauthorized Service Deregistration \cite{VelepuchaFlores2023Microservices}, \cite{Zirpins2020ESOCC}, \cite{Alshuqayran2016Systematic} \\
 &  & V$_{11}$: Replay Attacks on Service Requests \cite{Monteiro2018MicroserviceSecurity}, \cite{ECSA2022}, \cite{genfer2021avoiding} \\
 &  & V$_{12}$: Service Impersonation \cite{Chelladhurai2016SecuringDocker}, \cite{muresu2021investigating}, \cite{pereira2019security}, \cite{habbal2020enhancing}, \cite{Burns2018DesigningDistributedSystems} \\
\cline{2-3}
  & \multirow{6}{*}{\scriptsize Network Segmentation and Isolation} & V$_{13}$:  Poor Network Segmentation \cite{Esposito2016Microservices}, \cite{SurveyMSAAdvancedManufacturing} \\
 &  & V$_{14}$: Misconfigured Service Mesh \cite{SultanAhmadDimitriou2019}, \cite{DiFrancesco2019Microservices} \\
 &  & V$_{15}$: Misconfigured Network Access \cite{pereira2019security}, \cite{SurveyMSAAdvancedManufacturing} \\
 &  & V$_{16}$: Improper Firewall Configuration \cite{pereira2019security}, \cite{SurveyMSAAdvancedManufacturing} \\
 &  & V$_{17}$: Segmentation Bypass \cite{gkioulos2017security}, \cite{suomalainen2019defense} \\
 &  & V$_{18}$: Default Network Configurations \cite{dragoni2017microservices}, \cite{DiFrancesco2019Microservices} \\
\cline{2-3}
 & \multirow{6}{*}{\scriptsize  In-Transit Data Protection} & V$_{19}$: Weak In-Transit Data Encryption \cite{prabath2020}, \cite{Wolff2017Microservices} \\
 &  & V$_{20}$: Faulty Certificate Validation \cite{kramer2019implementing}, \cite{Barczak2021Performance} \\
 &  & V$_{21}$: In-Transit Metadata Exposure \cite{newman2015}, \cite{Wolff2017Microservices} \\
 &  & V$_{22}$: Encryption Key Mismanagement \cite{hannousse2021securing}, \cite{Barczak2021Performance} \\
 &  & V$_{23}$: Inadequate Network Certificate Validation \cite{Liu2020Microservices}, \cite{Wolff2017Microservices} \\
 &  & V$_{24}$: Hardcoded Encryption Keys \cite{Loureiro2021SecuringMicroservices}, \cite{Wolff2017Microservices} \\
\cline{2-3}
 & \multirow{5}{*}{\scriptsize Throttling and Rate Limiting} & V$_{25}$: Absence of Dynamic Rate Limiting \cite{muresu2021investigating}, \cite{Burns2018DesigningDistributedSystems}, \cite{Shabani2021ModernDistributedSystems} \\
 &  & V$_{26}$: Lack of IP-Based Rate Limiting \cite{Waseem2020}, \cite{Shabani2021ModernDistributedSystems} \\
 &  & V$_{27}$: Generic Rate Limiting Policies \cite{Shabani2021ModernDistributedSystems}, \cite{Burns2018DesigningDistributedSystems} \\
&  & V$_{28}$: Misconfigured Rate Limits \cite{kalubowila2021optimization, Shabani2021ModernDistributedSystems} \\
 &  & V$_{29}$: Token Bucket Overflow \cite{genfer2021avoiding}, \cite{Burns2018DesigningDistributedSystems} \\
\cline{1-3} 
\multirow{21}{*}{\small Data Consistency and Integrity} & \multirow{5}{*}{\scriptsize  Data Storage} & V$_{30}$: Weak Database Encryption \cite{MateusCoelho2021SecuringMicroservices}, \cite{Garriga2018Taxonomy} \\
 &  & V$_{31}$: Insufficient Database Hardening \cite{PreuveneersJoosen2017}, \cite{Soylemez2022FeatureDriven} \\
 &  & V$_{32}$: Exposure of Default Database Credentials \cite{Yu2017}, \cite{Garriga2018Taxonomy} \\
 &  & V$_{33}$: Sensitive Data in Error Messages \cite{esposito2017security}, \cite{Soylemez2022FeatureDriven} \\
 &  & V$_{34}$: Absence of Data Integrity Checks \cite{esposito2017security}, \cite{Garriga2018Taxonomy} \\
\cline{2-3}
 & \multirow{4}{*}{\scriptsize Data Sanitization and Validation } & V$_{35}$: SQL Injection \cite{Yarygina2018}, \cite{Soylemez2022FeatureDriven} \\
 &  &  V$_{36}$: Cross-Site Scripting \cite{Yarygina2018}, \cite{Soylemez2022FeatureDriven} \\
 &  & V$_{37}$: Command Injection \cite{Yarygina2018}, \cite{OWASP2023} \\
 &  & V$_{38}$: Insecure Data Deserialization \cite{Loureiro2021SecuringMicroservices}, \cite{Soylemez2022FeatureDriven} \\
\cline{2-3}
 & \multirow{4}{*}{\scriptsize Data Access Control} & V$_{39}$: Direct Object Reference \cite{SamoylenkoSelivanova2023}, \cite{Richards2015MicroservicesSOA} \\
 &  & V$_{40}$:  Excessive Privilege Granting \cite{FPS2018} \\
 &  & V$_{41}$: Hardcoded Credentials Exposure: \cite{BarrocaFilho2019MicroserviceHealth}, \cite{Richards2015MicroservicesSOA} \\
 &  & V$_{42}$: Privilege Escalation \cite{Driss2021MicroservicesIoT, nehme2019fine} \\
\cline{2-3}
& \multirow{4}{*}{\scriptsize Data Management and Consistency} & V$_{43}$: Race Condition Exploitation \cite{muresu2021investigating}, \cite{DiFrancesco2019Microservices} \\
 &  & V$_{44}$: Inadequate Data Transaction Management \cite{SamoylenkoSelivanova2023}, \cite{DiFrancesco2019Microservices} \\
 &  & V$_{45}$: Weak Data Synchronization \cite{MazzaraOberSalaun2018STAF} \\
 &  & V$_{46}$: Concurrent Data Access Mismanagement \cite{ECSA2022}, \cite{DiFrancesco2019Microservices} \\
\cline{2-3}
& \multirow{4}{*}{\scriptsize Data Backup and recovery} & V$_{47}$: Irregular Backups Maintenance \cite{Driss2021MicroservicesIoT}, \cite{Soylemez2022FeatureDriven} \\
 &  & V$_{48}$: Insecure Backup Storage \cite{Driss2021MicroservicesIoT}, \cite{Soylemez2022FeatureDriven} \\
 &  & V$_{49}$: Lack of Backup Validation \cite{Esposito2016Microservices}, \cite{Soylemez2022FeatureDriven} \\
 &  & V$_{50}$: Improper Backup Disposal \cite{gotz2018challenges}, \cite{Soylemez2022FeatureDriven} \\
\cline{1-3}
\multirow{20}{*}{\small Authentication and Authorization} & \multirow{4}{*}{\scriptsize Digital Identity Management} & V$_{51}$: Weak Authentication Protocols \cite{guija2018identity} \\
 &  & V$_{52}$: Account Enumeration Risks \cite{newman2015}, \cite{MicroservicesBuildingScalable} \\
 &  & V$_{53}$: Use of Compromised Credentials \cite{MicroservicesBuildingScalable} \\
 &  & V$_{54}$: Misconfigured Identity Federation \cite{MicroservicesBuildingScalable} \\
\cline{2-3}
 & \multirow{4}{*}{\scriptsize Permission and Privilege Management} & V$_{55}$:  Excessive Privilege Provisions \cite{Burns2018DesigningDistributedSystems}, \cite{Soylemez2022FeatureDriven} \\
 &  & V$_{56}$: Insufficient Token Invalidation \cite{jalonen2022security, Burns2018DesigningDistributedSystems, tai2023framework} \\
 &  & V$_{57}$: Insecure Token Storage \cite{Burns2018DesigningDistributedSystems}, \cite{Soylemez2022FeatureDriven} \\
 &  & V$_{58}$: Hardcoded Credentials \cite{ponce2022smells, Burns2018DesigningDistributedSystems} \\
\cline{2-3}
 & \multirow{4}{*}{\scriptsize Authentication Mechanisms} & V$_{59}$:  Password Reuse \cite{nehme2019fine, Richards2015MicroservicesSOA} \\
 &  & V$_{60}$: Insecure Password Recovery \cite{nehme2019fine, Richards2015MicroservicesSOA} \\
 &  & V$_{61}$: Lack of Multi-Factor Authentication \cite{rudrabhatla2020security, Richards2015MicroservicesSOA} \\
 &  & V$_{62}$: Phishing Vulnerability \cite{dohnalmicroservices, prabath2020} \\
\cline{2-3}
 & \multirow{4}{*}{\scriptsize Authorization and Policy Enforcement} & V$_{63}$: Unenforced Access Controls \cite{rudrabhatla2020security, Richards2015MicroservicesSOA} \\
 &  & V$_{64}$: Access Grant Misconfigurations \cite{ahmadvand2016requirements}, \cite{prabath2020} \\
 &  & V$_{65}$: Direct Object Exposure \cite{chaves2019systematic} \\
 &  & V$_{66}$: Over-privileged APIs \cite{Richards2015MicroservicesSOA} \\
\cline{2-3}
 & \multirow{4}{*}{\scriptsize Session Security} & V$_{67}$: Session Hijacking \cite{Burns2018DesigningDistributedSystems}, \cite{guija2018identity} \\
 &  & V$_{68}$: Cross-Site Request Forgery \cite{Burns2018DesigningDistributedSystems}, \cite{OWASP2023} \\
 &  & V$_{69}$: Session Control Exploitation \cite{Burns2018DesigningDistributedSystems}, \cite{guija2018identity} \\
 &  & V$_{70}$: Improper Session Expiry \cite{Burns2018DesigningDistributedSystems}, \cite{guija2018identity} \\
\cline{1-3}
\multirow{16}{*}{\small Deployment and Orchestration} & \multirow{4}{*}{\scriptsize Container Security} & V$_{71}$: Container Misconfigurations \cite{Loureiro2021SecuringMicroservices}, \cite{Garriga2018Taxonomy} \\
 &  & V$_{72}$: Improper Container Isolation \cite{MazzaraOberSalaun2018STAF}, \cite{Garriga2018Taxonomy} \\
 &  & V$_{73}$: Direct Storage of Sensitive Data on Container Image \cite{Liu2020Microservices}, \cite{Garriga2018Taxonomy} \\
 &  & V$_{74}$: Outdated/Insecure Container Image Usage \cite{Torkura2017}, \cite{Garriga2018Taxonomy} \\
\cline{2-3}
 & \multirow{4}{*}{\scriptsize Orchestration Platform} & V$_{75}$: Misconfigured Orchestration Dashboards \cite{Martin2018DockerVulnerability}, \cite{habbal2020enhancing} \\
 &  & V$_{76}$: Unrestricted Orchestration API Access \cite{SurveyMSAAdvancedManufacturing}, \cite{Yarygina2018} \\
 &  & V$_{77}$: Improper permissions \cite{Jander2018DefenseDepth}, \cite{huang2020learn} \\
 &  & V$_{78}$: Inherent Orchestration Tool Vulnerabilities \cite{Torkura2018, SurveyMSAAdvancedManufacturing, huang2020learn} \\
\cline{2-3}
 & \multirow{4}{*}{\scriptsize Configuration Management} & V$_{79}$: Insecure Service Configuration \cite{Martin2018DockerVulnerability}, \cite{Wolff2017Microservices} \\
 &  & V$_{80}$: Lack of Configuration Validation \cite{Martin2018DockerVulnerability}, \cite{Wolff2017Microservices} \\
 &  & V$_{81}$: Embedded Secrets in Configuration Files \cite{Martin2018DockerVulnerability}, \cite{Wolff2017Microservices} \\
 &  & V$_{82}$: Insecure Configuration File Validation \cite{Martin2018DockerVulnerability}, \cite{Wolff2017Microservices} \\
\cline{2-3}
 & \multirow{4}{*}{\scriptsize Dependency Management} & V$_{83}$: Integration of Vulnerable Components \cite{Wolff2017Microservices} \\
 &  & V$_{84}$: Use of Outdated Dependencies \cite{Beyah2018SecurityPrivacy}, \cite{Wolff2017Microservices} \\
 &  & V$_{85}$: Insufficient Dependency Scanning \cite{Waseem2020}, \cite{Wolff2017Microservices} \\
 &  & V$_{86}$: Lack of Transitive Dependency Validation \cite{Wolff2017Microservices} \\
\cline{1-3}
\multirow{20}{*}{\small Heterogeneous Features} & \multirow{4}{*}{\scriptsize Diversified Technology Stacks} & V$_{87}$: Inconsistent Security Practices \cite{rudrabhatla2020security} \\
 &  & V$_{88}$: Issues in Specific Libraries \cite{Barczak2021Performance}, \cite{dragoni2017microservices} \\
 &  & V$_{89}$:  Platform Misconfigurations \cite{Barczak2021Performance}, \cite{dragoni2017microservices} \\
 &  & V$_{90}$: Patch Management Complexity \cite{Barczak2021Performance} \\
\cline{2-3}
 & \multirow{4}{*}{\scriptsize Interoperability and Compatibility} & V$_{91}$: Legacy System Integration Vulnerabilities \cite{VelepuchaFlores2023Microservices}, \cite{Barczak2021Performance} \\
 &  & V$_{92}$: Mismatched Data Formats \cite{Barczak2021Performance} \\
 &  & V$_{93}$: Service Mesh Configuration Errors \cite{habbal2020enhancing}, \cite{Barczak2021Performance} \\
 &  & V$_{94}$: Inconsistent Integration Security \cite{gotz2018challenges}, \cite{Barczak2021Performance} \\
\cline{2-3}
 & \multirow{4}{*}{\scriptsize Third-Party Dependencies} & V$_{95}$: Compromised Supply Chains \cite{muresu2021investigating}, \cite{CarneiroSchmelmer2016Microservices} \\
 &  & V$_{96}$: Third-Party Service Outages \cite{Richards2015MicroservicesSOA}, \cite{CarneiroSchmelmer2016Microservices} \\
 &  & V$_{97}$: Insecure Third-Party Components \cite{MateusCoelho2021SecuringMicroservices}, \cite{CarneiroSchmelmer2016Microservices} \\
 &  & V$_{98}$: Poor Security Practices in Third-Party Components \cite{Barczak2021Performance}, \cite{CarneiroSchmelmer2016Microservices} \\
\cline{2-3}
 & \multirow{4}{*}{\scriptsize Framework and Language Vulnerabilities} & V$_{99}$: Injection Vulnerabilities \cite{newman2015}, \cite{Barczak2021Performance} \\
 &  & V$_{100}$: Improper XSS Preventions \cite{prabath2020}, \cite{newman2015}, \cite{Barczak2021Performance} \\
 &  & V$_{101}$: Insecure Deserialization \cite{newman2015}, \cite{Barczak2021Performance} \\
 &  & V$_{102}$: Frameworks-Specific Flaws \cite{newman2015}, \cite{Barczak2021Performance} \\
\cline{2-3}
 & \multirow{4}{*}{\scriptsize Platform and Infrastructure } & V$_{103}$: Cloud Misconfiguration \cite{Monteiro2018MicroserviceSecurity}, \cite{Tenev2020MicroserviceSecurity} \\
&  & V$_{104}$: Misconfigured Infrastructure Tools \cite{Belair2019ContainerSecurity}, \cite{Tenev2020MicroserviceSecurity} \\
 &  & V$_{105}$: VMs Mismanagement \cite{DiFrancesco2019Microservices}, \cite{Tenev2020MicroserviceSecurity} \\
 &  & V$_{106}$: Vulnerable Heterogeneous Networks \cite{prabath2020}, \cite{Tenev2020MicroserviceSecurity} \\
\cline{1-3}
\multirow{20}{*}{\small DevOps and CI/CD Practices} & \multirow{4}{*}{\scriptsize CI/CD Pipeline} & V$_{107}$: Vulnerable CI/CD Tools \cite{gotz2018challenges}, \cite{SurveyMSAAdvancedManufacturing} \\
 &  & V$_{108}$: CI/CD Misconfigurations \cite{singh2022securing, Balalaie2016MicroservicesDevOps} \\
 &  & V$_{109}$: Lack of Pipeline Security Controls \cite{SurveyMSAAdvancedManufacturing}, \cite{Baskarada2020Microservices} \\
 &  & V$_{110}$: Deploying Vulnerable Code \cite{MazzaraOberSalaun2018STAF, singh2022securing, SurveyMSAAdvancedManufacturing} \\
\cline{2-3}
 & \multirow{4}{*}{\scriptsize Infrastructure as Code} & V$_{111}$: Insecure IaC Scripts \cite{singh2022securing, NkomoCoetzee2019SecureMicroservices, SurveyMSAAdvancedManufacturing} \\
&  & V$_{112}$: Misconfigured Automated Deployment \cite{RademacherSorgallaSachweh2018}, \cite{Beyah2018SecurityPrivacy} \\
 &  & V$_{113}$: Hardcoded Secrets in IaC \cite{SurveyMSAAdvancedManufacturing} \\
 &  & V$_{114}$: No IaC Version Control \cite{SurveyMSAAdvancedManufacturing} \\
\cline{2-3}
 & \multirow{4}{*}{\scriptsize Secrets Management} & V$_{115}$: Poor Secrets Rotation \cite{Billawa2022MicroserviceSecurity}, \cite{VelepuchaFlores2023Microservices} \\
 &  & V$_{116}$: Lack of Centralized Secrets Management \cite{Billawa2022MicroserviceSecurity}, \cite{VelepuchaFlores2023Microservices} \\
 &  & V$_{117}$:  Secrets in Version Control \cite{Soylemez2022FeatureDriven}, \cite{Belair2019ContainerSecurity} \\
 &  & V$_{118}$: Unmonitored Secrets Access \cite{Billawa2022MicroserviceSecurity}, \cite{VelepuchaFlores2023Microservices} \\
\cline{2-3}
 & \multirow{4}{*}{\scriptsize Testing and Analysis} & V$_{119}$: Lack of Automated Testing \cite{prabath2020}, \cite{Soylemez2022FeatureDriven} \\
 &  & V$_{120}$: Poor Manual Security Reviews \cite{newman2015}, \cite{NkomoCoetzee2019SecureMicroservices}, \cite{Soylemez2022FeatureDriven} \\
 &  & V$_{121}$: Ignoring Test Results \cite{newman2015}, \cite{Soylemez2022FeatureDriven} \\
 &  & V$_{122}$: Outdated Testing Tools  \cite{prabath2020}, \cite{Soylemez2022FeatureDriven} \\
\cline{2-3}
 & \multirow{4}{*}{\scriptsize DevSecOps Integration} & V$_{123}$: Inadequate Security Training \cite{chandramouli2022implementation, SurveyMSAAdvancedManufacturing}, \cite{MateusCoelho2021SecuringMicroservices} \\
 &  & V$_{124}$: Neglecting Early Security \cite{lee2023microservices, mczara2020modeling, SurveyMSAAdvancedManufacturing}, \cite{MateusCoelho2021SecuringMicroservices} \\
&  & V$_{125}$: Inconsistent Security Practices  \cite{chandramouli2022implementation, SurveyMSAAdvancedManufacturing}, \cite{MateusCoelho2021SecuringMicroservices} \\
&  & V$_{126}$: Poor Security-DevOps Integration \cite{mczara2020modeling, SurveyMSAAdvancedManufacturing, MateusCoelho2021SecuringMicroservices} \\\hline
\end{longtable}
}
\twocolumn

\textit{V$_{1}$ - Exposed API Endpoints:} This vulnerability refers to the risk posed by API endpoints that lack adequate authentication mechanisms. It arises from insufficient implementation of authentication controls, enabling unauthorized access to sensitive data and application functionalities \cite{RademacherSorgallaSachweh2018}. Exploitation of this vulnerability compromises data integrity and allows unauthorized manipulation of critical application operations \cite{Garriga2018Taxonomy}. For example, CVE-2023-28346 demonstrates attackers exploiting this issue to gain unauthorized access and manipulate application requests \cite{mitre001}.

\vspace{0.03in}

\textit{V$_{2}$ - Accidental Exposure of Sensitive API Endpoints:} This vulnerability occurs when sensitive API endpoints are inadvertently exposed to unauthorized users or the public \cite{Garriga2018Taxonomy}. It arises from misconfigurations or oversights during application development \cite{prabath2020}. Exploiting this vulnerability results in unauthorized access, compromising the confidentiality and integrity of sensitive data and system functionalities \cite{CarneiroSchmelmer2016Microservices}. For instance, CVE-2019-11248 highlights the accidental exposure of Kubernetes dashboards, allowing unrestricted access to sensitive information and control panels \cite{mitre002}.

\vspace{0.03in}

\textit{V$_{3}$ - Untrusted Third-Party APIs:} This vulnerability refers to the risk posed by third-party APIs from unknown or untrusted sources within microservices-based applications \cite{prabath2020}. It arises from potential security threats such as injection attacks and data breaches exploiting these APIs. Exploitation of this vulnerability grants malicious actors unauthorized access to the system through compromised APIs \cite{CarneiroSchmelmer2016Microservices}. An example is CVE-2023-52077, where attackers executed system commands within an application by exploiting unknown and untrusted APIs \cite{mitre003}.

\vspace{0.03in}

\textit{V$_{4}$ - Weak API Authentication:} This vulnerability is characterized by inadequate authentication mechanisms implemented for APIs within microservice architectures \cite{prabath2020}. It arises due to insecure authentication protocols that fail to properly verify and authorize access requests \cite{PereiraVale2021}. Exploitation of this vulnerability leads to unauthorized access, allowing malicious actors to bypass authentication mechanisms with minimal effort \cite{Beyah2018SecurityPrivacy}. A significant example is CVE-2024-24771, where an attacker may use compromised super user credentials to bypass multi-factor authentication  \cite{mitre004}.

\vspace{0.03in}

\textit{V$_{5}$ - Insecure API Deserialization:} This vulnerability is characterized by improper handling or validation of serialized data within applications. Gateways often act as intermediaries that route and process API requests, including serialized data exchanges between services. When deserialization processes are insecure, attackers can exploit this weakness by sending malicious serialized data through the API endpoints. This can lead to remote code execution, unauthorized access, data manipulation, or replay attacks, as the malicious payload is deserialized and executed within the system \cite{Alshuqayran2016Systematic}. An instance of this vulnerability is CVE-2024-4200, where insecure serialization led to remote code execution on application servers \cite{mitre005}.

\vspace{0.03in}

\textit{V$_{6}$ - Misconfigured API Gateways:} This vulnerability occurs due to inadequate or incorrect security configurations in API gateways. It arises when the security settings of the API gateway are insufficient, exposing sensitive internal services to the public. This misconfiguration allows users to bypass security checks or gain unauthorized access to sensitive data and functions \cite{DeAlmeida2022}. CVE-2018-1002105 exemplifies this vulnerability, where misconfigured Kubernetes API servers enabled unauthorized requests to backend services, circumventing security measures \cite{mitre006}.

\vspace{0.03in}

\begin{tcolorbox}[left=0pt, top=0pt, right=0pt, bottom=0pt, boxrule=0.75pt]
Exploiting vulnerabilities in API Gateways and Endpoints leads to unauthorized access, data manipulation, and extensive network compromise in microservice architectures. This disrupts application continuity, compromises data integrity, and undermines overall network security. Robust security measures and continuous monitoring are essential to mitigate these risks.
\end{tcolorbox}

\subsubsection{Service Discovery}
Service discovery is a mechanism used in microservice architectures to automatically detect and manage the availability of services within a network \cite{Alshuqayran2016Systematic}. This process is essential for maintaining the dynamic nature of microservice architectures, enabling services to locate and interact with each other without manual intervention \cite{Wolff2017Microservices}. Features of Service Discovery include service registration, health checks, and service lookups. The security of services is vital; any compromise results in unauthorized modifications to service registrations, leading to significant information breaches and undermining the system's reliability \cite{VelepuchaFlores2023Microservices}. If service discovery is exploited, it could allow attackers to manipulate service traffic, leading to disruptions and unauthorized control over service interactions. In the following, we describe vulnerabilities inherent to Service Discovery mechanisms.

\textit{V$_{7}$ - Service Registration Tampering:} This vulnerability is characterized by the malicious modification or registration of services within the service discovery mechanism of a microservices architecture. It arises due to the attacker's ability to manipulate service traffic, gaining unauthorized control over service interactions \cite{BarrocaFilho2019MicroserviceHealth}. Exploitation of this vulnerability leads to significant consequences, such as service disruptions or unauthorized execution of operations, exploiting the relationship between service discovery and registration processes \cite{PriyaSethuramanKhan2023}. An example illustrating similar principles is the DNS poisoning attack described in CVE-2024-21661, where unauthorized manipulation of service registrations compromised system integrity \cite{mitre007}.

\vspace{0.03in}

\textit{V$_{8}$ - Unauthorized Service Discovery Access:} This vulnerability occurs when a threat actor gains unauthorized access to the service discovery system within the microservice architecture. It exposes sensitive information about the service architecture, allowing attackers to gather intelligence about the application's internal structure \cite{BarrocaFilho2019MicroserviceHealth}. The exploitation of this vulnerability enables various attacks, such as man-in-the-middle (MITM) or service impersonation attacks \cite{muresu2021investigating}. For example, CVE-2024-41701 highlights the consequences of an unauthorized access attempt \cite{mitre008}.

\vspace{0.03in}

\textit{V$_{9}$ - Service Registration Validation Gaps:} This vulnerability is characterized by the absence of proper or sufficient mechanisms within a microservice architecture application to validate newly registered services at the time of registration. It arises due to the system's failure to enforce validation checks, allowing malicious actors to register services with malicious intent or modify critical services undetected. Exploitation of this vulnerability results in service disruptions or unauthorized modifications \cite{PriyaSethuramanKhan2023}. While direct CVE links are rare due to their oversight nature, their impact conceptually aligns with incidents such as CVE-2021-44228, where inadequate validation compromised system security \cite{mitre009}.

\vspace{0.03in}

\textit{V$_{10}$ - Unauthorized Service Deregistration:} This vulnerability allows malicious actors to remove legitimate services from the service registry of a microservices application without proper authorization. It occurs due to insufficient controls preventing unauthorized deregistration of services, potentially impacting service availability within the application \cite{Zirpins2020ESOCC}. Exploitation of this vulnerability leads to severe consequences, such as service disruptions. While establishing a direct CVE link is challenging due to specific deployment environments, the vulnerability's impact mirrors incidents like CVE-2024-30381, where unauthorized actions threatened system integrity \cite{mitre0010}.

\vspace{0.03in}

\textit{V$_{11}$ - Replay Attacks on Service Requests:} This vulnerability involves attackers intercepting and replaying valid service requests within a microservices architecture. It exploits the lack of mechanisms to detect or prevent replayed requests, leading to unauthorized access to services or disclosure of sensitive information \cite{Monteiro2018MicroserviceSecurity}. A similar vulnerability is illustrated by CVE-2023-34625, where an attacker may intercept Bluetooth Low Energy (BLE) requests and later replay them to gain unauthorized access \cite{mitre0011}.

\vspace{0.03in}

\textit{V$_{12}$ - Service Impersonation:} This vulnerability occurs when an attacker impersonates a legitimate service within a microservice architecture application, deceiving other services into believing it is authentic. The vulnerability arises due to inadequate authentication or verification mechanisms, allowing unauthorized services to masquerade as legitimate ones \cite{Chelladhurai2016SecuringDocker}. Exploitation of this vulnerability leads to unauthorized data access or manipulation of sensitive information. An example with similar implications is CVE-2022-39388, which can lead to service impersonation \cite{mitre0012}.

\begin{tcolorbox}[left=0pt, top=0pt, right=0pt, bottom=0pt, boxrule=0.75pt]
Service discovery vulnerabilities allow attackers to manipulate the registration or discovery process, potentially leading to unauthorized access and service disruptions. This can result in data breaches and compromised system integrity. Mitigation measures include strong authentication, secure communications, and continuous monitoring of the service registry.
\end{tcolorbox}

\subsubsection{\textit{Network Segmentation and Isolation}}
Network Segmentation and Isolation refer to the strategy of dividing the network into distinct segments to enhance security within microservice architectures \cite{jin2020anomaly}. This approach minimizes the attack surface by preventing unauthorized access to critical services and data. Features include segmenting services based on functionality and implementing isolation mechanisms to control inter-service communication. The security of network segmentation is crucial as improper segmentation exposes sensitive services to broader network access, leading to potential breaches and unauthorized access. If network segmentation is compromised, attackers could move laterally within the network, accessing sensitive data and disrupting services \cite{SurveyMSAAdvancedManufacturing, DiFrancesco2019Microservices}. In this subsection, we will explore the vulnerabilities that arise from improper network segmentation, highlighting the critical importance of correctly implementing this strategy to enhance security.

\textit{V$_{13}$ - Poor Network Segmentation:} This vulnerability is defined by the absence or inadequate segmentation of the network within a microservices-based application. It arises due to insufficient isolation between services that should be segmented for security reasons. Exploitation of this vulnerability compromises the availability of microservices, as sensitive services become accessible from the broader network \cite{Esposito2016Microservices}. An example illustrating the risks associated with poor network segmentation is CVE-2023-6229, highlighting the importance of robust network segmentation to mitigate security risks \cite{mitre0013}.

\vspace{0.03in}

\textit{V$_{14}$ - Misconfigured Service Mesh:} This vulnerability occurs when there is an incorrect or suboptimal setup of a service mesh within a microservices architecture. It allows attackers to manipulate inter-service communication, potentially disrupting network segmentation processes \cite{SultanAhmadDimitriou2019}. For example, CVE-2024-40632 allows attackers to trigger a denial-of-service (DoS) attack, impacting service mesh stability and network segmentation \cite{mitre0014}.

\vspace{0.03in}

\textit{V$_{15}$ - Misconfigured Network Access:} This vulnerability is characterized by incorrect policies and settings governing network access controls. It arises due to misconfigurations that inadvertently permit access to the network. The exploitation of this vulnerability leads to unauthorized lateral movement within the network and access to sensitive functions during network segmentation and isolation processes \cite{pereira2019security}. An example illustrating this vulnerability is CVE-2024-25124, highlighting the risks associated with misconfigured network access controls \cite{mitre0015}.

\vspace{0.03in}

\textit{V$_{16}$ - Improper Firewall Configuration:} This vulnerability occurs due to the incorrect setup of firewall rules and policies, deviating from recommended or secure practices \cite{pereira2019security}. Such misconfigurations can result in inadequate separation of services, exposing sensitive data and enabling attackers to move laterally within the network. The consequences include potential data breaches and service disruptions. An example of this vulnerability is CVE-2024-3384, which illustrates the potential impact of improperly configured firewalls \cite{mitre0016}.

\vspace{0.03in}

\textit{V$_{17}$ - Segmentation Bypass:} Containers are often isolated from each other and the host system to maintain security and prevent unauthorized access. This vulnerability allows an attacker to bypass these isolation boundaries by exploiting the TOCTOU race condition, potentially gaining unauthorized access to the host filesystem or other containers \cite{gkioulos2017security}. An example illustrating the consequences of such vulnerabilities is CVE-2018-15664, highlighting the importance of ensuring secure interactions between different segments of a network, particularly when dealing with containerized environments \cite{mitre0017}.

\vspace{0.03in}

\textit{V$_{18}$ - Default Network Configurations:} This vulnerability refers to the use of default settings in network devices within a microservices architecture. Default configurations, including unchanged passwords and default port numbers, potentially allow attackers to gain unauthorized access to microservices while implementing network segmentation and isolation \cite{dragoni2017microservices}. Exploitation of this vulnerability leads to unauthorized control over network devices, disruption of network services, and increased susceptibility to further attacks. An example illustrating the risks associated with default configurations is CVE-2020-7802, highlighting the importance of securing network devices against such vulnerabilities \cite{mitre0018}.

\begin{tcolorbox}[left=0pt, top=0pt, right=0pt, bottom=0pt, boxrule=0.75pt]
Vulnerabilities such as inadequate network segmentation, misconfigured service meshes, and flawed access controls in microservice architectures pose significant risks. Addressing these vulnerabilities with robust security measures is crucial to prevent unauthorized access, data breaches, and ensure system integrity and availability.
\end{tcolorbox}

\subsubsection{\textit{In-Transit Data Protection}}
In-Transit Data Protection ensures the security of data as it moves between services in a microservices environment. This involves implementing robust encryption measures to prevent data interception and tampering during transmission. In-transit data protection features include secure communication protocols, encryption of data packets, and validation of communication channels \cite{Barczak2021Performance}. The security of in-transit data is imperative; vulnerabilities here lead to data breaches, exposing sensitive information and causing serious compliance issues. If in-transit data protection is compromised, attackers could intercept and manipulate data, compromising the confidentiality and integrity of the system \cite{Wolff2017Microservices}. Below is a categorized list of identified vulnerabilities in this domain, detailing their descriptions and impacts.

\textit{V$_{19}$ - Weak In-Transit Data Encryption:} This vulnerability arises when data transmitted between services is not properly encrypted. This can expose sensitive information to interception or tampering by malicious actors. Common causes include using outdated encryption protocols, weak cipher suites, or failing to encrypt data altogether. Such vulnerabilities can lead to data breaches, unauthorized access, or data manipulation. An example of this vulnerability is CVE-2016-2183, also known as the "SWEET32" vulnerability, which illustrates one manifestation of this issue \cite{mitre0019}.

\vspace{0.03in}

\textit{V$_{20}$ - Faulty Certificate Validation:} This vulnerability within microservices architecture highlights inadequate validation of network certificates used for inter-service communication \cite{Liu2020Microservices}. It arises due to insufficient validation procedures for network certificates, allowing potential exploitation by attackers to conduct MITM attacks. Exploitation of this vulnerability could lead to interception and manipulation of communications between services, causing operational disruptions and unauthorized control over service interactions. For instance, CVE-2024-28872 underscores the risks associated with improperly validated certificates, emphasizing the critical need for robust certificate validation practices to mitigate such vulnerabilities in a similiar context \cite{mitre0020}.

\vspace{0.03in}

\textit{V$_{21}$ - In-Transit Metadata Exposure:} This vulnerability occurs when sensitive metadata, such as headers, routing information, or other protocol-specific details, is unintentionally disclosed during data transmission between microservices \cite{newman2015}. This information can include details like IP addresses, service versions, or internal network configurations. If exposed, attackers can use this metadata to gather intelligence about the system's architecture, identify potential vulnerabilities, or map the network. This can lead to targeted attacks, such as service discovery, unauthorized access, or even denial-of-service attacks. For example, CVE-2018-7750 refers to an issue in Jenkins that exposed metadata, potentially leading to information disclosure \cite{mitre0021}.

\vspace{0.03in}

\textit{V$_{22}$ - Encryption Key Mismanagement:} This vulnerability refers to the improper handling of encryption keys used to secure data during transmission between microservices. This can include storing keys in insecure locations, failing to rotate keys regularly, or using weak or easily guessable keys. Such practices make it easier for attackers to gain access to these keys, enabling them to decrypt sensitive data being transmitted. Mismanagement can also occur if keys are shared insecurely between services or users, increasing the risk of unauthorized access. For example, CVE-2016-0733 pertains to OpenSSL and describes an issue where certain conditions could lead to the improper handling of cryptographic keys \cite{mitre0022}.

\vspace{0.03in}

\textit{V$_{23}$ - Inadequate Network Certificate Validation:} This vulnerability stems from inadequate validation of network certificates in microservices architecture, introducing vulnerabilities in inter-service communication. Exploitation of this vulnerability facilitates MITM attacks, where an attacker intercepts and manipulates communication, leading to operational disruptions and unauthorized control over service interactions \cite{Liu2020Microservices}. An example demonstrating this vulnerability is CVE-2024-25124, emphasizing the risks associated with improperly validated certificates \cite{mitre0023}.

\vspace{0.03in}

\textit{V$_{24}$ - Hardcoded Encryption Keys:} This vulnerability occurs when encryption keys are hardcoded directly into the source code of microservices, making them vulnerable to exploitation by malicious actors \cite{Loureiro2021SecuringMicroservices}. If attackers gain access to these hardcoded keys, they can decrypt sensitive data being transmitted, compromising the confidentiality and integrity of communications. Additionally, hardcoded keys cannot be easily rotated or updated, increasing the risk of long-term exposure. An example demonstrating the risks of hardcoding credentials is CVE-2024-25731, highlighting the security implications of such insecure practices \cite{mitre0024}.

\vspace{0.03in}

\begin{tcolorbox}[left=0pt, top=0pt, right=0pt, bottom=0pt, boxrule=0.75pt]
Effective in-transit data protection is crucial for securing microservices. Addressing vulnerabilities such as weak encryption, faulty certificate validation, and poor key management is vital. Ensuring the confidentiality and integrity of data during transmission safeguards the entire microservices environment from significant security risks.
\end{tcolorbox}

\subsubsection{\textit{Throttling and Rate Limiting}}
In microservice architectures, throttling and rate limiting are critical controls that manage the influx of service requests, ensuring that no single service is overwhelmed and resources are fairly allocated. Features within this category include dynamic rate limiting, IP-based rate limiting, and rate limiting policies. The security of throttling and rate limiting is paramount as ineffective controls expose services to denial-of-service attacks and performance bottlenecks. If compromised, attackers could flood the services with excessive requests, leading to system outages and degraded performance, ultimately affecting the availability and reliability of the microservices \cite{Shabani2021ModernDistributedSystems, Burns2018DesigningDistributedSystems}.

\textit{V$_{25}$ - Absence of Dynamic Rate Limiting:} This vulnerability is characterized by the absence of adequate rate limiting mechanisms within the microservices architecture, which fails to restrict the rate of incoming requests effectively. It arises due to oversight in implementing robust rate limiting across services. Exploitation of this vulnerability leads to Denial of Service (DoS) attacks, brute-force attacks, and resource exhaustion, severely impacting system availability. \cite{muresu2021investigating} discuss instances where the lack of proper rate limiting in microservices left systems vulnerable, highlighting the critical need for such measures. An example of this vulnerability is CVE-2023-48840, which demonstrates how proper rate limiting could have mitigated attacks, underscoring its importance in defending against brute-force attempts \cite{mitre0025}.

\vspace{0.03in}

\textit{V$_{26}$ - Lack of IP-Based Rate Limiting:} This vulnerability manifests when there are no tailored rate limiting policies applied to different services, user behaviors, or IP addresses within the microservices architecture. It is caused by the absence of customized rate limiting strategies, leaving the system vulnerable to various attacks such as DoS, DDoS, or brute-force attacks. The impact of this vulnerability is profound, affecting system availability and performance under stress conditions. \cite{Shabani2021ModernDistributedSystems} identifies vulnerabilities in microservices arising from the lack of individualized rate limiting policies, highlighting the critical need for adaptive and scalable rate limiting mechanisms. An example illustrating this vulnerability includes CVE-2018-6389, which underscores the repercussions of inadequate rate limiting strategies in mitigating similar attack vectors, emphasizing the necessity of tailored defenses in microservice environments \cite{mitre0026}.

\vspace{0.03in}

\textit{V$_{27}$ - Generic Rate Limiting Policies:} This vulnerability occurs within the microservices architecture when there are no tailored rate-limiting policies applied to different services, user behavior, or IP addresses \cite{Shabani2021ModernDistributedSystems}. It arises due to the oversight of not implementing customized rate limiting strategies. Exploitation of this vulnerability may lead to attacks such as DoS, DDoS, or brute-force attacks. \cite{Shabani2021ModernDistributedSystems} discusses vulnerabilities arising from the lack of individualized rate limiting, stressing the importance of adaptive policies. For instance, CVE-2023-20155 provides an illustrative example of how similar vulnerabilities could be exploited, emphasizing the need for tailored defenses in microservice environments \cite{mitre0027}.

\vspace{0.03in}

\textit{V$_{28}$ - Misconfigured Rate Limits:} This vulnerability refers to errors in setting up rate-limiting mechanisms that are either too strict, too lenient, or inconsistently applied within microservice architectures \cite{Shabani2021ModernDistributedSystems}. It occurs during initial configurations, posing operational challenges and increasing the risk of DoS attacks. Consequently, these misconfigurations lead to cascading failures, impacting the availability and reliability of the system. \cite{Shabani2021ModernDistributedSystems} highlights the risks associated with improperly configured rate limits, emphasizing the importance of precise settings to mitigate vulnerabilities. An example illustrating this vulnerability is CVE-2023-20155, which highlights the repercussions of inadequate rate limiting configurations in mitigating potential attacks, emphasizing the critical need for accurate and effective rate limiting measures \cite{mitre0028}.

\vspace{0.03in}

\textit{V$_{29}$ - Token Bucket Overflow:} This vulnerability arises when a token bucket algorithm mismanages token accumulation and consumption. If tokens accumulate excessively or are not properly consumed, attackers can exploit this to send a large burst of requests, potentially overwhelming the system. This can lead to resource exhaustion or denial-of-service conditions \cite{genfer2021avoiding}. An example illustrating this vulnerability is CVE-2024-0555, which demonstrates how vulnerabilities in API components can be exploited, highlighting the critical importance of securing APIs against malicious exploitation to safeguard sensitive data and maintain system reliability \cite{mitre0029}
\vspace{0.03in}

\begin{tcolorbox}[left=0pt, top=0pt, right=0pt, bottom=0pt, boxrule=0.75pt]
Effective throttling and rate limiting are essential for maintaining the stability and security of microservices. Vulnerabilities such as the absence of dynamic rate limiting, lack of IP-based rate limiting, generic rate limiting policies, misconfigured rate limits, and token bucket overflow pose significant risks. Properly managing these controls ensures the system's availability and reliability, preventing denial of service attacks and performance bottlenecks.
\end{tcolorbox}

\subsection{\textbf{Data Consistency and Integrity}}
In microservice architectures, data consistency and integrity ensure that data across independent databases remains accurate, reliable, and accessible. This practice is essential for the stability and reliability of microservice architectures, as inconsistencies or data corruption lead to erroneous outputs, malfunctioning services, and a loss of user trust \cite{newman2015}. If data consistency and integrity are compromised, it disrupts operations significantly and potentially leads to data breaches, compromising an entire application's integrity \cite{Garriga2018Taxonomy}.

\subsubsection{Data Storage}
Data storage in microservices involves secure storage and access to data collected by various services, ensuring integrity and confidentiality. This subcategory is critical; vulnerabilities such as weak encryption or insufficient database hardening pose significant risks to external threats \cite{Soylemez2022FeatureDriven}. If these vulnerabilities are exploited, it leads to data breaches, unauthorized access, and compliance challenges, causing reputational damage and compromising the system's confidentiality and integrity \cite{Garriga2018Taxonomy}.

\textit{V$_{30}$ - Weak Database Encryption:} This vulnerability within the microservices architecture occurs when data collected by different services within the application is stored in databases with inadequate or absent encryption protocols \cite{MateusCoelho2021SecuringMicroservices}. It is caused by the failure to implement strong encryption practices for stored data. Exploitation of this vulnerability may lead to data breaches and unauthorized access. For instance, CVE-2023-4420 highlights the importance of robust database encryption to prevent such security risks \cite{mitre0030}.

\vspace{0.03in}

\textit{V$_{31}$ - Insufficient Database Hardening:} This vulnerability refers to insufficient security measures and configurations within databases used by various services within microservices architectures \cite{PreuveneersJoosen2017}. It occurs due to the lack of rigorous hardening practices. The impact includes potential data leaks and unauthorized access, which compromise data confidentiality and integrity. For example, CVE-2023-2986 exemplifies the risks associated with inadequate database hardening efforts \cite{mitre0031}.

\vspace{0.03in}

\textit{V$_{32}$ - Exposure of Default Database Credentials:} This critical vulnerability arises within the microservices architecture when default database credentials are not changed before deployment \cite{Yu2017}. It occurs due to oversight in updating default configurations. Exploitation of this vulnerability allows easy unauthorized access to databases, compromising the confidentiality and integrity of stored data. \cite{Yu2017} highlights the risks associated with using default database credentials, emphasizing the importance of implementing strong authentication practices. For example, CVE-2024-22432 illustrates similar risks associated with default configurations in other contexts \cite{ mitre0032}.

\vspace{0.03in}

\textit{V$_{33}$ - Sensitive Data in Error Messages:} This vulnerability involves exposing confidential information in error messages related to data storage \cite{esposito2017security}. It is caused by improper handling of error messages that fail to obfuscate sensitive data. This can reveal database structure, such as table and column names, or even actual sensitive data like personal information or financial details. For example, CVE-2023-27860 highlights the dangers of such exposures in other contexts \cite{mitre0033}.

\vspace{0.03in}

\textit{V$_{34}$ - Absence of Data Integrity Checks:} This vulnerability refers to the absence of mechanisms to ensure that stored data remains accurate and consistent \cite{esposito2017security}. Without such verification, data may become corrupted during storage or transmission, and unauthorized modifications may go undetected. This can result in unreliable data, system errors, and potential security breaches. For example, CVE-2020-0601 provides insights into similar risks associated with inadequate integrity checks in different scenarios \cite{mitre034}.

\vspace{0.03in}

\begin{tcolorbox}[left=0pt, top=0pt, right=0pt, bottom=0pt, boxrule=0.75pt]
The data storage vulnerabilities highlighted include weak encryption, inadequate security hardening, default credentials, exposure of sensitive information in error messages, and lack of data integrity checks. Together, these issues create a comprehensive risk profile, exposing systems to unauthorized access, data breaches, and potential manipulation. Addressing these vulnerabilities requires a holistic approach to secure database systems, ensuring robust encryption, proper configuration, access control, and consistent data verification to maintain data integrity and confidentiality.
\end{tcolorbox}

\subsubsection{\textit{Data Sanitization and Validation}}
In microservice architectures, Data Sanitization and Validation involve cleaning and verifying user input to ensure it is safe and conforms to expected formats. This process protects against security vulnerabilities like SQL injection and cross-site scripting, ensuring the integrity and security of data exchanged between services \cite{Soylemez2022FeatureDriven}.  Exploitation of vulnerabilities related to data sanitization and validation leads to unauthorized access, data breaches, and significant disruptions to the system's confidentiality, integrity, and availability. Ensuring robust data sanitization and validation is crucial to prevent security breaches that exploit unsanitized or unvalidated data. In the following, we discuss vulnerabilities related to data sanitization and validation.

\textit{V$_{35}$ - SQL Injection:} This vulnerability occurs when an attacker exploits improper handling of user inputs \cite{Yarygina2018}. By injecting malicious SQL commands into input fields, they can manipulate database queries, leading to unauthorized data access, modification, or deletion. This vulnerability typically arises from the lack of input validation and parameterized queries. In microservices, where multiple components interact with various databases, SQL injection can have cascading effects across the system. For example, CVE-2024-33551 highlights similar threats posed by SQL injection in other contexts \cite{mitre0035}.

\vspace{0.03in}

\textit{V$_{36}$ - Cross-Site Scripting:} Cross-Site Scripting (XSS) involves injecting malicious scripts into a web application, affecting users who interact with the service \cite{Yarygina2018}. In microservice architectures, XSS can compromise the security of interconnected services by allowing attackers to steal cookies, session tokens, or other sensitive information from users. It can also enable attackers to impersonate users and perform unauthorized actions. This vulnerability often arises from inadequate input and output encoding. For instance, CVE-2024-5385 highlights the risks and impacts associated with XSS attacks \cite{mitre0036}.

\vspace{0.03in}

\textit{V$_{37}$ - Command Injection:} This vulnerability occurs when an application improperly handles untrusted input, allowing an attacker to execute arbitrary commands on the host operating system \cite{OWASP2023}. This can lead to full system compromise, data theft, or destruction. This vulnerability typically results from failing to sanitize and validate inputs that are passed to system commands. In a microservices environment, command injection can impact multiple services if the compromised service has access to shared resources. For example, CVE-2019-12735 highlights specific instances where command injection vulnerabilities were exploited, underscoring the critical importance of robust command handling mechanisms \cite{mitre0037}.

\vspace{0.03in}

\textit{V$_{38}$ - Insecure Data Deserialization:} This vulnerability occurs when applications deserialize untrusted data without proper validation \cite{Loureiro2021SecuringMicroservices}. This can lead to remote code execution, data tampering, or other malicious actions. In microservices, where data is frequently exchanged between services, insecure deserialization can compromise multiple components. This vulnerability arises when data objects are serialized and deserialized without verifying their integrity or authenticity. CVE-2019-2725 exemplifies the risks posed by insecure deserialization in different scenarios, emphasizing the need for secure deserialization practices in microservice environments \cite{mitre0038}.

\vspace{0.03in}

\begin{tcolorbox}[left=0pt, top=0pt, right=0pt, bottom=0pt, boxrule=0.75pt]
Robust data sanitization and validation are essential to secure microservices. Vulnerabilities like SQL injection, XSS, command injection, and insecure data deserialization pose serious threats. Mitigating these vulnerabilities is crucial to safeguard against unauthorized access, data breaches, and maintain the overall security and stability of the system.
\end{tcolorbox}

\subsubsection{\textit{Data Access Control}}
Data access control within microservices architecture involves regulating who can access data within the system, ensuring that only authorized entities have access to sensitive information. Robust data access controls are essential to safeguard privacy and maintain system integrity, preventing unauthorized access and data leaks. Weak access controls expose the system to significant vulnerabilities from both internal and external threats \cite{Richards2015MicroservicesSOA}. This subsection explores vulnerabilities related to data access control, discussing the potential dangers of inadequate controls.

\textit{V$_{39}$ - Direct Object Reference Vulnerability:} This vulnerability allows user input to directly reference objects within the application's architecture, posing a significant threat to systems utilizing microservices architecture \cite{SamoylenkoSelivanova2023}. It arises due to insufficient validation or sanitization of user inputs, enabling attackers to potentially access unauthorized data and perform unintended operations. Exploitation of this vulnerability lead to unauthorized data access, data modification, and potential disruption of application functionality. For example, scenarios akin to CVE-2024-5128 illustrate the impact of such vulnerabilities in this context \cite{mitre0039}.

\vspace{0.03in}

\textit{V$_{40}$ - Excessive Privilege Granting:} This vulnerability occurs when applications within microservices architecture grant higher privileges than necessary, expanding the attack surface and enabling unauthorized actions. It violates the principle of least privilege, leading to unauthorized access to sensitive data, system resource manipulation, and potential service disruptions. Exploitation of this vulnerability result in significant security breaches, as demonstrated by CVE-2024-29210, which illustrates the consequences of similar vulnerabilities in other scenarios \cite{mitre0040}.

\vspace{0.03in}

\textit{V$_{41}$ - Hardcoded Credentials Exposure:} This vulnerability arises when developers embed credentials directly into the source code of microservices architecture applications \cite{BarrocaFilho2019MicroserviceHealth}. It enables potential unauthorized access to system resources through hardcoded credentials, leading to unauthorized system access and data breaches. Exploitation of this vulnerability severely impact system security, as seen in CVE-2024-0865, which highlights the risks associated with hardcoded credentials in other contexts \cite{mitre0041}.

\vspace{0.03in}

\textit{V$_{42}$ - Privilege Escalation:} This vulnerability emerges when attackers exploit improperly configured permissions within microservice architectures \cite{Driss2021MicroservicesIoT}. It allows unauthorized access to sensitive data or functionalities, enabling attackers to gain higher privileges than intended. Exploitation of this vulnerability lead to unauthorized data modifications, disruptions in service, and unauthorized access to critical system components. CVE-2024-2338 provides insights into similar risks associated with privilege escalation in different scenarios \cite{mitre0042}.

\vspace{0.03in}

\begin{tcolorbox}[left=0pt, top=0pt, right=0pt, bottom=0pt, boxrule=0.75pt]
Strong data access control is essential in microservices architecture to safeguard against unauthorized access and data breaches. Key vulnerabilities, including direct object references, excessive privilege granting, hardcoded credentials, and privilege escalation, can compromise system security. Mitigating these issues is critical for ensuring the confidentiality, integrity, and reliability of the system, as highlighted by real-world CVE examples.
\end{tcolorbox}

\subsubsection{\textit{Data Management and Consistency}}
In microservice architectures, maintaining data consistency and integrity ensures that data across various independent data stores remains accurate and reliable, which is crucial as each service may manage its own unique data store. This practice is essential for the stability and reliability of the system. Inconsistencies or data corruption lead to erroneous outputs, malfunctioning services, and a loss of user trust. Vulnerabilities related to data management and integrity disrupt operations significantly, potentially leading to data breaches \cite{DiFrancesco2019Microservices}.

\textit{V$_{43}$ - Race Condition Exploits:} This vulnerability arises in applications utilizing microservices architecture, where attackers exploit timing discrepancies between requests and responses to manipulate race conditions \cite{muresu2021investigating}. This manipulation leads to unpredictable outcomes and compromises the actions of services, potentially resulting in data corruption or unauthorized access. For instance, CVE-2024-5558 demonstrates general exploits involving race conditions in other contexts \cite{mitre0043}.

\vspace{0.03in}

\textit{V$_{44}$ - Inadequate Data Transaction Management:} This vulnerability refers to inadequate handling of data consistency and rollback mechanisms within microservice architectures \cite{SamoylenkoSelivanova2023}. Insufficient transaction management leads to data loss or inconsistencies across services, impacting the reliability of the system. Exploitation of this vulnerability may result in incomplete or incorrect data transactions, as illustrated by CVE-2023-34362, which highlights the impact of similar vulnerabilities in different scenarios \cite{mitre0044}.

\vspace{0.03in}

\textit{V$_{45}$ - Weak Data Synchronization:} This vulnerability exposes weaknesses in data synchronization mechanisms across services within microservice architectures \cite{MazzaraOberSalaun2018STAF}. It may lead to unauthorized access, data leaks, or data corruption. Exploitation of insecure data synchronization results in compromised data integrity and unauthorized access to sensitive information. For example, CVE-2024-6778 underscores the importance of secure data synchronization practices in mitigating such risks \cite{mitre0045}.

\vspace{0.03in}

\textit{V$_{46}$ - Concurrent Data Access Mismanagement:} This vulnerability occurs due to improper management of concurrent data access within microservices architecture \cite{ECSA2022}. Mismanagement leads to data corruption, inconsistencies, or unintended behaviors across services, affecting the integrity of shared data between different components. Exploitation of this vulnerability may lead to data integrity violations or operational disruptions. For example, CVE-2024-2032 illustrates risks associated with mismanaged concurrent data access in other contexts \cite{mitre0046}.

\vspace{0.03in}

\begin{tcolorbox}[left=0pt, top=0pt, right=0pt, bottom=0pt, boxrule=0.75pt]
Ensuring data consistency and integrity in microservices is crucial for system reliability. Vulnerabilities such as race condition exploits, inadequate transaction management, weak synchronization, and concurrent data access mismanagement lead to data corruption, inconsistencies, and unauthorized access. Addressing these issues is essential to ensure stable and accurate data across all services.\end{tcolorbox}

\subsubsection{\textit{Data Backup and Recovery}}
Data Backup and Recovery strategies in microservices are essential to ensure that data can be restored promptly and accurately in the event of a data breach or system failure. It includes features such as regular backup maintenance, secure storage, validation of backups, and proper disposal of obsolete backups. Robust backup and recovery strategies are crucial for maintaining the integrity and continuity of the microservices architecture. If these vulnerabilities are exploited, they result in data loss, unauthorized access, and system downtime, compromising data integrity and confidentiality \cite{Soylemez2022FeatureDriven}. Below is a list of vulnerabilities categorized with their descriptions and impacts.
  
\textit{V$_{47}$ - Irregular Backup Maintenance:} This vulnerability exposes applications to significant risks due to the absence of regular backups, leaving the system vulnerable to unexpected failures such as cyber-attacks, human errors, or equipment malfunctions \cite{Driss2021MicroservicesIoT}. It arises from neglecting to establish and maintain a proper backup strategy. Exploitation of this vulnerability leads to data loss, prolonged downtime, and inability to restore critical services, severely impacting the system's availability. For example, CVE-2023-6271 illustrates the consequences of insufficient backup practices in other contexts \cite{mitre0047}.

\vspace{0.03in}

\textit{V$_{48}$ - Insecure Backup Storage:} This vulnerability occurs within microservices architecture when data and configuration backups are stored insecurely, allowing unauthorized access and tampering by malicious actors \cite{Driss2021MicroservicesIoT}. It arises due to inadequate security measures in storing backups. Exploitation of this vulnerability results in unauthorized access to sensitive data, data tampering, and potential data breaches, compromising the integrity and confidentiality of the stored information. For example, CVE-2023-5808 highlights the risks associated with insecure data storage practices \cite{mitre0048}.

\vspace{0.03in}

\textit{V$_{49}$ - Lack of Backup Validation:} This vulnerability arises when backups of data and configurations are not regularly validated for integrity, recoverability, and completeness within microservices architecture \cite{Esposito2016Microservices}. It occurs due to the absence of robust validation processes for backups. Exploitation of this vulnerability results in the inability to restore data correctly, leading to incomplete or corrupted data restoration efforts during recovery scenarios. CVE-2023-6553 underscores the importance of regularly validating backups in mitigating risks \cite{mitre0049}.

\vspace{0.03in}

\textit{V$_{50}$ - Improper Backup Disposal:} This vulnerability occurs when outdated backups containing sensitive data and configurations are not properly disposed of at the end of their retention period within microservices architecture \cite{gotz2018challenges}. It arises due to inadequate procedures for the disposal of obsolete backups. Exploitation of this vulnerability results in unauthorized access to sensitive information through improperly discarded backups. For example, CVE-2024-20358 emphasizes the critical need for proper data disposal practices to mitigate potential risks \cite{mitre0050}.

\vspace{0.03in}

\begin{tcolorbox}[left=0pt, top=0pt, right=0pt, bottom=0pt, boxrule=0.75pt]
Robust backup and recovery strategies are critical in microservices architecture to prevent data loss, unauthorized access, and system downtime. Addressing vulnerabilities like irregular backup maintenance, insecure storage, lack of validation, and improper disposal is essential. These measures ensure data integrity and confidentiality, safeguarding the system's reliability and security. \end{tcolorbox}

\subsection{\textbf{Authentication and Authorization}}
Authentication and Authorization involve managing identity and access to ensure that only authenticated and authorized entities can access and control services and data within a microservices-based system. Authentication and Authorization security are paramount; a breach leads to unauthorized access, data breaches, and overall system compromise, significantly undermining the integrity and reliability of the microservices environment \cite{MicroservicesBuildingScalable} \cite{Soylemez2022FeatureDriven}.

\subsubsection{\textit{Digital Identity Management}}
Digital Identity Management in the microservices context explores the assignment of digital identities and access rights within the architecture \cite{Soylemez2022FeatureDriven}. It underscores the critical need for meticulous identity and access management to safeguard the system's integrity and security. If digital identity management security is compromised, it leads to unauthorized access, data breaches, and compromises the system's confidentiality, integrity, and availability \cite{MicroservicesBuildingScalable}. Below is a categorized overview of vulnerabilities identified in this domain, including their descriptions and impacts.

\textit{V$_{51}$ - Weak Authentication Protocols:} This vulnerability occurs in microservice architecture when authentication mechanisms lack robustness, failing to adequately protect against unauthorized system access \cite{guija2018identity}. It arises due to inadequate authentication protocols or weak password policies. Exploitation of this vulnerability leads to unauthorized access, system compromise, and breaches of system confidentiality and integrity. The Heartbleed bug (CVE-2014-0160) serves as a notable example illustrating the risks associated with such weaknesses \cite{mitre0051}.

\vspace{0.03in}

\textit{V$_{52}$ - Account Enumeration Risks:} This vulnerability arises when attackers exploit error messages or system behaviors in microservice architectures to identify service endpoints or user accounts \cite{newman2015}. It facilitates targeted exploits, credential stuffing, and brute-force attacks by revealing system structure information. Exploitation of this vulnerability leads to unauthorized access to sensitive data and system compromise. CVE-2021-29659 provides insights into the potential impacts of such vulnerabilities in a similar context \cite{mitre0052}.

\vspace{0.03in}

\textit{V$_{53}$ - Use of Compromised Credentials:} This vulnerability persists within microservice architecture systems when compromised credentials from past breaches continue to be used by users or services \cite{MicroservicesBuildingScalable}. It compromises system security by allowing attackers easy access using known compromised credentials. For instance, CVE-2024-5967 demonstrates the severe consequences of similar vulnerabilities in VPN services \cite{mitre0053}.

\vspace{0.03in}

\textit{V$_{54}$ - Misconfigured Identity Federation:} This vulnerability arises in microservice architectures due to misconfigured or improperly implemented identity federation protocols or settings between services \cite{MicroservicesBuildingScalable}. Misconfiguration leads to unauthorized system access, data leakage, and system compromise. CVE-2024-21643 illustrates the impact of misconfigured identity federation settings on Identity and Access Management (IAM) systems in a similar context \cite{mitre0054}.

\begin{tcolorbox}[left=0pt, top=0pt, right=0pt, bottom=0pt, boxrule=0.75pt]
Strong identity and access management is critical in microservices architecture to prevent unauthorized access and ensure data security. Addressing vulnerabilities such as weak authentication protocols, account enumeration, compromised credentials, and misconfigured identity federation is essential for maintaining system confidentiality, integrity, and availability.
\end{tcolorbox}

\subsubsection{\textit{Permission and Privilege Management}}
Permission and Privilege Management in microservices delves into the vulnerabilities that arise from overly lax permission and privilege management policies, which may grant excessive rights beyond what is necessary. It highlights the importance of implementing strict and well-defined permission and privilege management practices to ensure that each entity within the microservices architecture has only the access rights required for its function, thereby minimizing potential security risks \cite{Burns2018DesigningDistributedSystems}. Effective management of permissions and privileges is essential to address vulnerabilities like excessive privileges, inadequate token invalidation, insecure storage of access tokens, and embedded static credentials. If these vulnerabilities are exploited, it lead to unauthorized access, compromising system confidentiality and integrity.

\textit{V$_{55}$ - Excessive Privilege Provisions:} This vulnerability occurs within microservice architectures when entities are granted excessive privileges beyond what is necessary for their functions \cite{Burns2018DesigningDistributedSystems}. It stems from misconfigured access control policies or roles. Exploitation of this vulnerability lead to unauthorized access to sensitive information or services, compromising system confidentiality and integrity. CVE-2024-24830 illustrates the risk of such privilege escalation and its impact on system security \cite{mitre0055}.

\vspace{0.03in}

\textit{V$_{56}$ - Insufficient Token Invalidation:} In the microservice environment, this vulnerability arises when tokens used for entity access are not properly invalidated after use or session expiration \cite{Burns2018DesigningDistributedSystems}. This oversight allows attackers to reuse or steal these tokens for unauthorized system access, compromising confidentiality and integrity. CVE-2024-27782 provides insights into how such token misuse can be exploited by attackers \cite{mitre0056}.

\vspace{0.03in}

\textit{V$_{57}$ - Insecure Token Storage:} This vulnerability occurs when access tokens are stored insecurely within the microservice architecture, enabling attackers to exploit these tokens for unauthorized system access \cite{Burns2018DesigningDistributedSystems}. It compromises system security by allowing unauthorized parties to obtain and misuse these tokens. CVE-2024-20395 highlights the risks associated with insecure data storage practices \cite{mitre0057}.

\vspace{0.03in}

\textit{V$_{58}$ - Hardcoded Credentials:} This vulnerability arises when developers embed sensitive information such as login credentials and API keys directly into microservices' source code \cite{Burns2018DesigningDistributedSystems}. Attackers extract these credentials, potentially gaining unauthorized access and compromising system confidentiality and integrity. CVE-2024-39208, demonstrates the impact of similar issues on framework security \cite{mitre0058}.

\vspace{0.03in}

\begin{tcolorbox}[left=0pt, top=0pt, right=0pt, bottom=0pt, boxrule=0.75pt]
Effective permission and privilege management in microservices architecture is crucial to prevent unauthorized access and ensure system security. Key vulnerabilities include excessive privilege provisions, insufficient token invalidation, insecure token storage, and hardcoded credentials. Addressing these issues is essential to protect sensitive data and maintain system reliability. Real-world examples, such as various CVEs, highlight the significant risks posed by poor permission and privilege management and underscore the need for strict access control policies.\end{tcolorbox}

\subsubsection{\textit{Authentication Mechanisms}}
Authentication Mechanisms in the microservices context investigate the vulnerabilities that arise from improper implementation of identity and access authentication. Ensuring that the identities of entities within the microservices environment are properly and securely verified is crucial to prevent unauthorized access and potential security breaches. Robust authentication mechanisms are vital to maintaining the integrity and security of the microservices architecture \cite{Richards2015MicroservicesSOA}. If the security of authentication mechanisms is compromised, it leads to risks such as password reuse, insecure password recovery processes, lack of multi-factor authentication, and susceptibility to phishing attacks, ultimately resulting in unauthorized access and data breaches.

\textit{V$_{59}$ - Password Reuse:} Password reuse in microservice architectures occurs when users employ the same password across multiple accounts and services \cite{Richards2015MicroservicesSOA}. This practice significantly increases the risk that if one account is compromised, attackers use the same credentials to access other services. This leads to cascading security failures, exposing multiple systems to unauthorized access, data breaches, and potential system-wide compromises. The impact of password reuse on system security is highlighted by CVE-2023-38489 \cite{mitre0059}.

\vspace{0.03in}

\textit{V$_{60}$ - Insecure Password Recovery:} Insecure password recovery processes in microservice architectures lack adequate security measures \cite{Richards2015MicroservicesSOA}. Attackers exploit these weak recovery mechanisms to hijack the password reset process. This results in unauthorized system access, leading to data breaches and compromise of system integrity and security. CVE-2023-7028 provides insights into the risks associated with insecure password recovery processes \cite{mitre0060}.

\vspace{0.03in}

\textit{V$_{61}$ - Lack of Multi-Factor Authentication:} This vulnerability exists when microservice architectures do not implement multi-factor authentication (MFA) \cite{Richards2015MicroservicesSOA}. Without MFA, attackers need only compromise a single authentication factor, such as a password, to gain unauthorized access to the system. This significantly increases the risk of unauthorized access, leading to potential data breaches and system compromises. The importance of MFA is underscored by the consequences highlighted in CVE-2019-11510 \cite{mitre0061}.

\vspace{0.03in}

\textit{V$_{62}$ - Phishing Vulnerability:} Phishing vulnerabilities arise when attackers use social engineering techniques to deceive users into revealing sensitive information within microservice architectures \cite{prabath2020}. Phishing attacks exploit user trust to gain access to confidential data and system credentials \cite{goel2018mobile}. This leads to significant data leaks, unauthorized access, and compromise of system integrity and security. The severe impact of phishing attacks is illustrated by CVE-2019-16759 \cite{mitre0062}.

\begin{tcolorbox}[left=0pt, top=0pt, right=0pt, bottom=0pt, boxrule=0.75pt]
Effective authentication mechanisms in microservices are vital for security. Addressing specific vulnerabilities such as password reuse, insecure recovery processes, lack of MFA, and phishing attacks is critical. These weaknesses lead to unauthorized access and data breaches, underscoring the need for robust authentication practices to protect sensitive data and maintain system integrity. \end{tcolorbox}

\subsubsection{\textit{Authorization and Policy Enforcement}}

Authorization and policy enforcement examine the vulnerabilities stemming from inadequate authorization and access control policies, which lead to inconsistencies and security risks throughout the microservices environment \cite{Richards2015MicroservicesSOA}. Emphasizing the importance of stringent authorization mechanisms and consistent policy enforcement, this section underscores the need for comprehensive access control strategies to safeguard the integrity and security of the system.

\vspace{0.03in}

\textit{V$_{63}$ - Unenforced Access Controls:} This vulnerability arises in microservice architectures when proper access control mechanisms are not implemented or enforced, allowing unauthorized users to access system resources \cite{Richards2015MicroservicesSOA}. This oversight leads to unauthorized data access, manipulation, or disclosure, potentially disrupting system operations. CVE-2021-44877 exemplifies the severe consequences of inadequate access controls \cite{mitre063}.

\vspace{0.03in}

\textit{V$_{64}$ - Access Grant Misconfigurations:} This vulnerability occurs when developers or administrators within microservice architectures incorrectly assign excessive system access privileges \cite{prabath2020}. Such misconfigurations lead to unauthorized data access, potential breaches, or inappropriate activities within the system, significantly impacting system operations. CVE-2021-41635 illustrates the significant impact of human errors in access control management \cite{mitre064}.

\vspace{0.03in}

\textit{V$_{65}$ - Direct Object Exposure:} In microservice architectures, this vulnerability arises when applications expose internal objects to external entities without proper validation or authorization checks \cite{chaves2019systematic}. Exploitation of this vulnerability allows attackers to access data or manipulate system functionalities, potentially leading to significant operational disruptions. CVE-2024-5166 provides insights into the risks associated with Insecure Direct Object Reference (IDOR) vulnerabilities \cite{mitre065}.

\vspace{0.03in}

\textit{V$_{66}$ - Over-privileged APIs:} This vulnerability exists in microservice architectures where APIs managing sensitive data are overly privileged or inadequately secured \cite{Richards2015MicroservicesSOA}. Attackers exploit these vulnerabilities to gain unauthorized access, manipulate data, or disrupt operations, significantly impacting system functionality. CVE-2021-40444 underscores the consequences of inadequate API access controls \cite{mitre066}.

\begin{tcolorbox}[left=0pt, top=0pt, right=0pt, bottom=0pt, boxrule=0.75pt]
Robust authorization and policy enforcement are crucial in microservices architecture to prevent vulnerabilities like unenforced access controls, misconfigured access grants, direct object exposure, and over-privileged APIs. These issues lead to unauthorized access and data breaches, compromising system security. Relevant CVEs highlight the need for stringent measures to protect the integrity and confidentiality of microservices environments.
\end{tcolorbox}

\subsubsection{\textit{Session Security}}

Session security in microservices investigates the vulnerabilities associated with improper handling and securing of sessions and tokens within the architecture \cite{Burns2018DesigningDistributedSystems}. Proper session management is critical to prevent unauthorized access and ensure that user sessions are maintained securely. Effective session security practices are vital for maintaining overall system security. If session security is compromised, it leads to unauthorized access, data breaches, and loss of user trust.

\vspace{0.03in}

\textit{V$_{67}$ - Session Hijacking:} This vulnerability manifests in microservice architectures when attackers seize active session tokens to impersonate users and gain unauthorized system access \cite{Burns2018DesigningDistributedSystems}. Such exploitation leads to significant security breaches involving sensitive user information. An illustrative example of this vulnerability is CVE-2017-3506, which underscores the severe implications of session management flaws \cite{mitre067}.

\vspace{0.03in}

\textit{V$_{68}$ - Cross-Site Request Forgery:} Within microservice architectures, this vulnerability occurs when attackers trick users into performing unintended actions within authenticated systems, potentially leading to unauthorized operations and data breaches \cite{OWASP2023}. For example, CVE-2021-22899 demonstrates the exploitation of similar vulnerabilities in web applications \cite{mitre068}.

\vspace{0.03in}

\textit{V$_{69}$ - Session Control Exploitation:} This vulnerability arises in microservice architectures when attackers exploit weaknesses in session management to gain unauthorized access to user sessions, facilitating session fixation or replay attacks \cite{guija2018identity}. For example, CVE-2022-39801 provides an example of such session management vulnerabilities affecting web applications \cite{mitre069}.

\vspace{0.03in}

\textit{V$_{70}$ - Improper Session Expiry:} This vulnerability occurs in microservice architectures when user sessions are not properly invalidated upon completion or expiration, allowing attackers to gain unauthorized access and potentially leading to security breaches \cite{Burns2018DesigningDistributedSystems}. For example, CVE-2019-16782 illustrates the implications of similar session expiry issues in web applications \cite{mitre070}.

\begin{tcolorbox}[left=0pt, top=0pt, right=0pt, bottom=0pt, boxrule=0.75pt]
Effective session management in microservice architectures is crucial for mitigating vulnerabilities such as session hijacking, cross-site request forgery, session control exploitation, and improper session expiry. These issues allow attackers to impersonate users and gain unauthorized access, compromising system security. \end{tcolorbox}

\subsection{\textbf{Deployment and Orchestration}}
Deployment and orchestration of microservices are critical components of their overall architecture, focusing on the tools, processes, and practices used during deployment, management, and scalability \cite{CarneiroSchmelmer2016Microservices}. This area examines the security concerns associated with containerization, orchestration platforms, and the operational challenges in service configuration management and lifecycle management. Proper deployment and management practices are essential to ensure the reliability and security of microservices. If these aspects are compromised, it leads to widespread vulnerabilities, system outages, and data breaches.

\subsubsection{\textit{Container  Security}}
Containerization is fundamental to the microservices architecture, making its security a vital consideration during system development \cite{Garriga2018Taxonomy}. Container security involves ensuring the secure configuration and operation of containers, which encapsulate microservice instances. It includes features such as secure container configuration, proper isolation, and secure storage practices. If container security is compromised, attackers gain unauthorized access, leading to potential data breaches and disruptions in service.

\textit{V$_{71}$ - Container Misconfigurations:} This vulnerability occurs due to incorrect setup of containers, such as Docker containers, within microservice architectures. Improper configuration settings create security gaps exploitable by malicious actors. Exploitation leads to unauthorized access to the containers and compromises the entire infrastructure \cite{Loureiro2021SecuringMicroservices}. An example is described in CVE-2019-5736, detailing the potential impact of misconfigurations on Docker systems \cite{mitre071}.

\vspace{0.03in}

\textit{V$_{72}$ - Improper Container Isolation:} This vulnerability arises from inadequate isolation between containers or between containers and the host system in microservice architectures. Insufficient isolation mechanisms allow attackers to gain unauthorized access to the infrastructure, execute unauthorized code, and disrupt service availability \cite{MazzaraOberSalaun2018STAF}. For example, CVE-2020-15257 highlights issues of poor container isolation \cite{mitre072}.

\vspace{0.03in}

\textit{V$_{73}$ - Direct Storage of Sensitive Data on Container Image:} This vulnerability arises when sensitive information, like credentials, is included directly within the container image in microservice architectures. Developers embedding sensitive data in container images expose it to potential attackers. Exploitation leads to unauthorized access to the image and extraction of sensitive information \cite{Liu2020Microservices}. For example, CVE-2023-38275 discusses the breach of sensitive information from container images \cite{mitre073}.

\vspace{0.03in}

\textit{V$_{74}$ - Outdated/Insecure Container Image Usage:} This vulnerability arises from deploying container images with known vulnerabilities or outdated dependencies within microservice architectures. Using insecure or obsolete container images lead to unauthorized access, potentially compromising the entire microservice infrastructure and causing security incidents such as service disruptions and data breaches \cite{Torkura2017}. For example, CVE-2018-14620 sheds light on the consequences of utilizing insecure container images \cite{mitre074}.

\begin{tcolorbox}[left=0pt, top=0pt, right=0pt, bottom=0pt, boxrule=0.75pt]
Ensuring robust deployment and orchestration in microservices is vital for system security and reliability. Key vulnerabilities, such as container misconfigurations, improper isolation, storing sensitive data in images, and using outdated or insecure images, pose significant risks. These issues lead to unauthorized access and compromise the entire infrastructure. Addressing these vulnerabilities is essential to enhance security, as highlighted by various CVEs. Implementing stringent security measures during containerization and deployment safeguards the integrity and efficiency of microservices environments.
\end{tcolorbox}

\subsubsection{\textit{Orchestration Platform}}
Various platforms and tools, such as Kubernetes, are employed for the management and deployment of service instances within the microservices architecture \cite{SurveyMSAAdvancedManufacturing}. It investigates the vulnerabilities that may arise from improper configuration of these orchestration tools and platforms.

\textit{V$_{75}$ - Misconfigured Orchestration Dashboards:} Various platforms and tools, such as Kubernetes, are employed for the management and deployment of service instances within the microservices architecture \cite{SurveyMSAAdvancedManufacturing}. Improper configuration of these orchestration tools and platforms leads to significant security vulnerabilities. If the orchestration platform's security is compromised, it results in unauthorized access, data breaches, and disruption of services. For example, CVE-2024-36819 signifies the impact of misconfigured dashboards \cite{mitre075}.

\vspace{0.03in}

\textit{V$_{76}$ - Unrestricted Orchestration API Access:} This vulnerability arises from insecure and unrestricted access to orchestration tool APIs within the microservice architecture. Developers or administrators failing to correctly secure and restrict access to these APIs create exploitable security gaps. Attackers gain unauthorized access and control over the infrastructure, leading to service disruptions and loss of data \cite{Yarygina2018}. An example is CVE-2013-6426, which illustrates the impact of insecure Orchestration API \cite{mitre076}.

\vspace{0.03in}

\textit{V$_{77}$ - Improper permissions:} This vulnerability occurs when roles and permissions are incorrectly or incompletely defined within microservice architectures. Developers or administrators who do not accurately define these roles and permissions expose the system to unauthorized access. This results in access to sensitive information and services, impacting the system's functionality \cite{Jander2018DefenseDepth}. For example, CVE-2024-4146 discusses the impact of Improper permissions \cite{mitre077}.

\vspace{0.03in}

\textit{V$_{78}$ - Inherent Orchestration Tool Vulnerabilities:} This vulnerability involves the presence of inherent flaws within the orchestration tools used in microservice architectures \cite{Torkura2018}. These flaws can be exploited by attackers to gain unauthorized access to the infrastructure, compromising the system's functionality. An example is CVE-2021-25742, which highlights the impact of vulnerabilities affecting the Kubernetes platform \cite{mitre078}.

\begin{tcolorbox}[left=0pt, top=0pt, right=0pt, bottom=0pt, boxrule=0.75pt]
Ensuring the secure configuration of orchestration platforms like Kubernetes is vital for maintaining the integrity of microservices architecture. Misconfigured dashboards, unrestricted API access, Improper permissions, and inherent tool vulnerabilities expose systems to unauthorized access, data breaches, and service disruptions. Mitigating these risks is essential for protecting the system's security and functionality. 
\end{tcolorbox}

\subsubsection{\textit{Configuration Management}}
Configuration Management in microservices examines the vulnerabilities that arise from improper or insecure management of various configurations within each service, as well as the handling of secrets \cite{Wolff2017Microservices}. This includes issues such as incorrect configuration settings, inadequate validation processes, and poor management of sensitive information. These vulnerabilities lead to significant security risks, including unauthorized access, data breaches, and compromised system integrity.

\textit{V$_{79}$ - Insecure Service Configuration:} This vulnerability arises from the improper configuration of services within the microservice architecture \cite{Martin2018DockerVulnerability}. Incorrect setup of services creates exploitable weaknesses, allowing attackers to compromise system operations. Exploitation leads to unauthorized access and control over the services. An example is CVE-2022-20308, which demonstrates the impact of insecure configurations \cite{mitre079}.

\vspace{0.03in}

\textit{V$_{80}$ - Lack of Configuration Validation:} This vulnerability occurs when service configurations are not properly validated before deployment within the microservice architecture. The absence of proper validation processes results in potential service disruptions and security risks. An example is highlighted in CVE-2020-16898, which underscores the consequences of misconfigurations due to the lack of proper validation \cite{mitre080} \cite{Martin2018DockerVulnerability}.

\vspace{0.03in}

\textit{V$_{81}$ - Embedded Secrets in Configuration Files:} This vulnerability arises when developers hardcode sensitive information, such as credentials and access tokens, directly into configuration files within microservice architectures. Such practices expose sensitive data to potential attackers. Exploitation leads to unauthorized access to sensitive information and services \cite{Martin2018DockerVulnerability}. An example is discussed in CVE-2023-33372, which explores the implications of hardcoded credentials \cite{mitre081}.

\vspace{0.03in}

\textit{V$_{82}$ - Insecure Configuration File Validation:} This vulnerability occurs due to inadequate validation of configuration files within the microservice architecture. Improper validation leads to attackers exploiting weaknesses within the files to manipulate system settings, compromising system operations \cite{Martin2018DockerVulnerability}. An example is described in CVE-2007-1741, which showcases the impact of insecure configuration file validation \cite{mitre082}.

\begin{tcolorbox}[left=0pt, top=0pt, right=0pt, bottom=0pt, boxrule=0.75pt]
Proper configuration management is vital in microservices architecture to mitigate security risks. Vulnerabilities like insecure service setups, insufficient validation of configurations, embedding sensitive data in configuration files, and poor validation practices lead to severe security breaches. Resolving these issues is crucial to maintain system integrity and prevent unauthorized access.\end{tcolorbox}

\subsubsection{\textit{Dependency Management}}
Dependency Management in microservices involves overseeing the updates and integrations of various service dependencies to ensure all components are secure and up-to-date. This aspect is crucial for maintaining system integrity and security. Proper management practices include integrating secure components, using updated libraries, and conducting thorough scanning of dependencies to mitigate security risks. Failure to manage dependencies effectively introduces vulnerabilities, leading to significant security breaches \cite{Wolff2017Microservices}. Ensuring robust dependency management is essential to prevent potential exploits and maintain a secure microservices environment.

\textit{V$_{83}$ - Integration of Vulnerable Components:} This vulnerability arises from the integration of external dependencies or frameworks with known vulnerabilities into the microservice architecture. Developers or administrators may not adequately assess the security risks of these components, leading to potential exploitation. Attackers gain unauthorized system access, impacting system operations \cite{Wolff2017Microservices}. An example is illustrated in CVE-2023-38295, which highlights the consequences of incorporating components with known vulnerabilities \cite{mitre083}.

\vspace{0.03in}

\textit{V$_{84}$ - Use of Outdated Dependencies:} This vulnerability occurs when dependencies are not regularly updated or maintained within the microservice architecture. Developers may neglect to keep dependencies up to date, leading to security risks. Exploitation results in unauthorized system access, service disruptions, and data breaches \cite{Beyah2018SecurityPrivacy}. For example, CVE-2024-3234 addresses the impact of using outdated and unmaintained dependencies \cite{mitre084}.

\vspace{0.03in}

\textit{V$_{85}$ - Insufficient Dependency Scanning:} This vulnerability arises from the neglect of comprehensive scans of libraries and dependencies within the microservice architecture. Administrators and developers may overlook weaknesses in dependencies, leading to potential security risks. Exploitation results in unauthorized system access and compromised system operations  \cite{Waseem2020}. For example, CVE-2019-15591 discusses the ramifications of inadequate dependency scanning \cite{mitre0085}.

\vspace{0.03in}

\textit{V$_{86}$ - Lack of Transitive Dependency Validation:} This vulnerability occurs when there is a lack of proper validation of indirect dependencies within the microservice architecture. Administrators and developers may not ensure the security of transitive dependencies, leading to potential security risks \cite{Wolff2017Microservices}. Exploitation results in unauthorized system access. For example, CVE-2013-6623 elucidates the impact of neglecting the security validation of vulnerable transitive dependencies \cite{mitre0086}.

\begin{tcolorbox}[left=0pt, top=0pt, right=0pt, bottom=0pt, boxrule=0.75pt]
Effective dependency management is essential for securing microservices architecture. Issues such as integrating vulnerable components, using outdated libraries, insufficient dependency scanning, and neglecting transitive dependency validation lead to severe security risks. These vulnerabilities provide attackers with opportunities to exploit system weaknesses and gain unauthorized access. Proper management practices are vital to maintaining system integrity and preventing security breaches.\end{tcolorbox}

\subsection{\textbf{Heterogeneous Features}}
In microservice architectures, heterogeneity refers to the utilization of a diverse array of programming languages, technologies, and frameworks within a single architecture. This diversity allows for unparalleled scalability and flexibility in program design, enabling the use of the best tools for specific tasks \cite{Barczak2021Performance}. However, this heterogeneity also introduces significant security challenges, as managing consistent security practices across varied components can be complex \cite{SurveyMSAAdvancedManufacturing}. The importance of securing a heterogeneous microservices environment cannot be overstated, as vulnerabilities arise from inconsistencies and integration issues, potentially leading to unauthorized access, data breaches, or system disruptions.

\subsubsection{\textit{Diversified Technology Stacks}}
Within microservices, diversified technology stacks are common, with different services often developed using various programming languages and technologies. This subcategory highlights the vulnerabilities associated with managing these diverse technologies \cite{Barczak2021Performance}. The presence of inconsistent security practices, issues in specific third-party libraries, platform misconfigurations, and the complexity of patch management are key concerns. If these vulnerabilities are exploited, they lead to unauthorized access, data breaches, and service disruptions, emphasizing the need for consistent security measures and effective management practices.

\textit{V$_{87}$ - Inconsistent Security Practices:} This vulnerability arises from the presence of inconsistent security practices across different technologies, microservices, or components within microservice architectures. Each component may employ varied security practices, standards, or configurations, creating gaps or weaknesses. Exploitation leads to attackers finding and exploiting these gaps, compromising the system. An example is illustrated in CVE-2021-47542, which demonstrates the consequences of inconsistent security configurations \cite{mitre0087}.

\vspace{0.03in}

\textit{V$_{88}$ - Issues in Specific Libraries:} This vulnerability arises from security risks due to issues in specific third-party libraries used within microservices. Outdated versions, lack of patching, or inherent design flaws in these libraries create security vulnerabilities. Exploitation leads to security breaches or unauthorized access. An example is highlighted in CVE-2017-5638, which sheds light on the impacts of using flawed libraries \cite{mitre088} \cite{dragoni2017microservices}.

\vspace{0.03in}

\textit{V$_{89}$ - Platform Misconfigurations:} This vulnerability occurs due to misconfigurations across various platforms within microservice architectures, such as cloud services, container orchestration tools, or databases. Insecure defaults, overly permissive access controls, or unimplemented encryption lead to security vulnerabilities. Exploitation results in unauthorized access, data breaches, or service disruptions. An example is discussed in CVE-2020-5902, which relates to the implications of misconfigurations \cite{mitre089} \cite{dragoni2017microservices}.

\vspace{0.03in}

\textit{V$_{90}$ - Patch Management Complexity:} This vulnerability arises from the complexity of patch management within the decentralized nature of microservices. Coordinating and applying patches across multiple services and components be challenging. Delayed patching of critical vulnerabilities increases the risk of exploitation by attackers. An example is described in CVE-2017-0144, which highlights the dangers of unpatched systems or delayed patching \cite{mitre090} \cite{Barczak2021Performance}.

\begin{tcolorbox}[left=0pt, top=0pt, right=0pt, bottom=0pt, boxrule=0.75pt]
The diverse technology stacks in microservices architecture offer flexibility but introduce security challenges. Inconsistent security practices, flawed third-party libraries, platform misconfigurations, and complex patch management lead to unauthorized access and data breaches. Consistent security measures and effective management practices are essential to maintaining a secure and reliable microservices environment.
\end{tcolorbox}

\subsubsection{Interoperability and Compatibility}
Interoperability and compatibility in microservices refer to the seamless integration and functioning of various frameworks and technologies within the architecture. Ensuring this interoperability is crucial, as integration challenges lead to security vulnerabilities and system inefficiencies \cite{Barczak2021Performance}. Vulnerabilities such as legacy system integration issues, mismatched data formats, service mesh configuration errors, and inconsistent integration security practices result in unauthorized access, data breaches, and service disruptions. Addressing these vulnerabilities through careful integration and consistent security measures is vital for maintaining a secure and efficient microservices environment.

\textit{V$_{91}$ - Legacy System Integration Vulnerabilities:} This vulnerability occurs when integrating legacy systems containing outdated or unsupported components and software within microservice architectures. The inherent security weaknesses in these older systems can be exploited by attackers, leading to unauthorized access and compromising the entire system's operations \cite{VelepuchaFlores2023Microservices}. An example is illustrated in CVE-2018-5547, which offers insight into the impact of such integration vulnerabilities \cite{mitre091}.

\vspace{0.03in}

\textit{V$_{92}$ - Mismatched Data Formats:} This vulnerability arises from the use of incompatible data formats for communication or data exchange among microservices or components within microservice architectures. Mismatched data formats across various technologies lead to data corruption, integrity loss, or parsing errors, facilitating injection attacks or data leakage\cite{Barczak2021Performance}. An example is highlighted in CVE-2015-4475, which underscores the consequences of data format inconsistencies \cite{mitre092}.

\vspace{0.03in}

\textit{V$_{93}$ - Service Mesh Configuration Errors:} This vulnerability occurs due to the misconfiguration of service mesh components within microservice architectures. Insecure default settings, inadequate access controls, or improper routing rules lead to security vulnerabilities. Exploitation results in unauthorized access, data leakage, or service disruptions. An example is discussed in CVE-2020-8559, which underscores the potential impacts of such misconfigurations \cite{mitre093} \cite{habbal2020enhancing}.

\vspace{0.03in}

\textit{V$_{94}$ - Inconsistent Integration Security:} This vulnerability arises from variations in security measures or protocols at integration points within microservice architectures. The use of different technologies leads to inconsistencies where microservices interact or exchange data. Exploitation results in unauthorized access, data breaches, or injection attacks \cite{gotz2018challenges}. An example is illustrated in CVE-2024-41656, which shows the effects this could have, in a similar context \cite{mitre094}.

\begin{tcolorbox}[left=0pt, top=0pt, right=0pt, bottom=0pt, boxrule=0.75pt]
Ensuring interoperability and compatibility in microservice architectures is essential to maintaining system security and efficiency. Vulnerabilities such as legacy system integration, mismatched data formats, service mesh configuration errors, and inconsistent integration security lead to unauthorized access, data breaches, and service disruptions. Addressing these vulnerabilities through careful integration and consistent security measures is crucial for a secure and reliable microservices environment.
\end{tcolorbox}

\subsubsection{Third-Party Dependencies}
In microservice architectures, integrating third-party components and services is often necessary for enhancing functionality or facilitating communication \cite{CarneiroSchmelmer2016Microservices}. It examines the vulnerabilities arising from reliance on these external elements. Key vulnerabilities include compromised supply chains, third-party service outages, insecure third-party components, and poor security practices in third-party components. Exploitation of these vulnerabilities leads to unauthorized access, data breaches, and service disruptions, highlighting the importance of stringent security measures and thorough vetting of third-party dependencies.

\textit{V$_{95}$ - Compromised Supply Chains:} This vulnerability arises when attackers compromise the software supply chain to inject malicious code or tamper with dependencies within the microservice architecture. Introducing compromised components, which contain vulnerabilities or backdoors, leads to unauthorized access, data theft, or service disruptions \cite{muresu2021investigating}. An example is outlined in CVE-2024-23332, which demonstrates the potential impact of a compromised supply chain \cite{mitre095}.

\vspace{0.03in}

\textit{V$_{96}$ - Third-Party Service Outages:} This vulnerability involves operational risks within microservice architectures due to reliance on third-party dependencies or services that may experience downtime. Decreased performance, disruptions, or unavailability of these services affect system functionality and availability \cite{Richards2015MicroservicesSOA}. An example is highlighted in CVE-2024-21815, which underscores the risks associated with dependency on third-party services \cite{mitre096}.

\vspace{0.03in}

\textit{V$_{97}$ - Insecure Third-Party Components:} This vulnerability arises from utilizing third-party libraries, frameworks, or services with security flaws within microservice architectures. Outdated versions or known security weaknesses in these components lead to threats like data breaches or remote code execution \cite{MateusCoelho2021SecuringMicroservices}. An example is discussed in CVE-2024-3454, which highlights the consequences of using insecure third-party components \cite{mitre097}.

\vspace{0.03in}

\textit{V$_{98}$ - Poor Security Practices in Third-Party Components:} This vulnerability is present when microservices depend on third-party components that lack robust security measures or adhere to insecure coding practices. Issues such as hardcoded credentials or insufficient input validation lead to attackers compromising system security \cite{Barczak2021Performance}. An example is illustrated in CVE-2023-7224, which shows the impact of poor security practices in third-party components \cite{mitre098}.

\begin{tcolorbox}[left=0pt, top=0pt, right=0pt, bottom=0pt, boxrule=0.75pt]
Integrating third-party dependencies in microservices architectures introduces significant security risks, such as compromised supply chains, service outages, insecure components, and poor security practices. These vulnerabilities lead to unauthorized access, data breaches, and service disruptions. Managing third-party dependencies with stringent security measures and thorough vetting is essential to mitigate these risks.
\end{tcolorbox}

\subsubsection{Framework and Language Vulnerabilities}
Microservices architecture often involves multiple teams working with different programming languages and frameworks. This diversity fosters innovation and flexibility but also introduces a range of inherent vulnerabilities \cite{newman2015}. Each framework and language has its own set of vulnerabilities, such as injection vulnerabilities and improper cross-site scripting prevention \cite{Barczak2021Performance}. These vulnerabilities significantly impact the overall security and performance of the system, leading to unauthorized access, data breaches, or system compromise. Understanding and addressing these specific vulnerabilities are essential for maintaining the security and reliability of a microservices environment.

\textit{V$_{99}$ - Injection Vulnerabilities:} This vulnerability arises due to the use of multiple languages and frameworks \cite{newman2015}. Each technology may have different methods and standards for input validation and handling. These inconsistencies can create system weaknesses, making it possible for attackers to exploit these gaps. As a result, microservices may become susceptible to various injection attacks, such as SQL injection, command injection, or script injection, leading to unauthorized access, data breaches, or system compromise. An example is CVE-2024-41944, which demonstrates the potential impact of injection attacks \cite{mitre099}.

\vspace{0.03in}

\textit{V$_{100}$ - Improper XSS Preventions:} This vulnerability occurs in microservice architectures due to improper implementation of XSS prevention measures. Insufficient input validation, lack of output encoding, or insecure coding practices lead to attackers injecting malicious scripts into microservice web-based applications. This enables arbitrary code execution in users' browsers, compromising data, or hijacking sessions. An example is CVE-2024-37165, which highlights the impact of XSS vulnerabilities within web applications \cite{mitre0100}.

\vspace{0.03in}

\textit{V$_{101}$ - Insecure Deserialization:} This vulnerability is a security risk within microservice architectures due to insecure deserialization processes, arising from the heterogeneity of technologies and frameworks. Attackers exploit inconsistencies in deserialization processes to execute code, manipulate data, or gain unauthorized access \cite{Barczak2021Performance}. This leads to critical issues in environments where data interchange is common. An example is CVE-2024-6152, which illustrates the consequences of insecure deserialization \cite{mitre0101}.

\vspace{0.03in}

\textit{V$_{102}$ - Framework-Specific Flaws:} This vulnerability concerns security flaws within the frameworks utilized across multiple services in microservice architectures. Weaknesses in the design, implementation, or configuration of these frameworks create exploitable gaps \cite{newman2015}. Exploitation leads to widespread security breaches. An example is CVE-2024-21312, which sheds light on the risks associated with flawed frameworks \cite{mitre0102}.

\begin{tcolorbox}[left=0pt, top=0pt, right=0pt, bottom=0pt, boxrule=0.75pt]
The use of diverse programming languages and frameworks introduces several vulnerabilities. These include injection attacks, improper XSS prevention, insecure deserialization, and framework-specific flaws. Each technology comes with its own set of security challenges, which significantly impact the overall system's security and performance. Addressing these vulnerabilities is crucial to maintaining a secure microservices environment. Real-world examples, such as various CVEs, highlight the importance of robust security practices to mitigate these risks effectively.
\end{tcolorbox}

\subsubsection{Platform and Infrastructure}

In microservice architectures, the Platform and Infrastructure refer to the essential technologies and systems that support the deployment, operation, and management of microservices \cite{Tenev2020MicroserviceSecurity}. This includes cloud environments, virtual machines (VMs), container orchestration platforms, and networking infrastructure. Key features within this category include cloud configuration, infrastructure tools, VMs management, and network security. The security of this category is crucial; if compromised, it leads to unauthorized access, data breaches, and significant service disruptions.

\textit{V$_{103}$ - Cloud Misconfiguration:} This vulnerability arises within microservice architectures due to misconfigurations in cloud environments, potentially exposing sensitive data, permitting unauthorized access, or creating exploitable weaknesses. These issues occur when cloud services, including IaaS, PaaS, and SaaS, are improperly configured or left with default or insecure settings \cite{Monteiro2018MicroserviceSecurity}. Exploitation leads to significant security incidents. An example is CVE-2023-31997, which highlights the consequences of misconfigured cloud resources \cite{mitre0103}.

\vspace{0.03in}

\textit{V$_{104}$ - Misconfigured Infrastructure Tools:} This vulnerability arises within microservice architectures due to misconfigured infrastructure tools, which are crucial for managing cloud resources, containers, networking, and more \cite{Belair2019ContainerSecurity}. When these tools are not configured according to security best practices, they become targets for attackers. Exploitation leads to severe security breaches. An example is CVE-2019-9946, which illustrates the impact of such misconfigurations \cite{mitre0104}.

\vspace{0.03in}

\textit{V$_{105}$ - VMs Mismanagement:} This vulnerability arises within microservice architectures due to the mismanagement of VMs, which host services and applications. Issues occur when VMs are improperly managed or secured, leading to unauthorized access, data breaches, and resource exploitation \cite{DiFrancesco2019Microservices}. Exploitation causes major security problems. An example is CVE-2024-30247, which demonstrates the risks associated with insecure or misconfigured VMs \cite{mitre0105}.

\vspace{0.03in}

\textit{V$_{106}$ - Vulnerable Heterogeneous Networks:} This vulnerability arises within microservice architectures due to inadequate network security for supporting infrastructure and platforms, exposing data, enabling unauthorized access, and facilitating attacks \cite{prabath2020}. Insufficient security controls such as segmentation, encryption, and monitoring make networks susceptible to exploitation. Exploitation of this vulnerability results in critical security incidents. An example is CVE-2022-4390, which outlines the dangers of network misconfigurations and vulnerabilities \cite{mitre0106}.

\begin{tcolorbox}[left=0pt, top=0pt, right=0pt, bottom=0pt, boxrule=0.75pt]
The heterogeneous nature of microservices architecture introduces several platform and infrastructure vulnerabilities. Misconfigurations in cloud environments, infrastructure tools, and virtual machines, as well as vulnerable heterogeneous networks, expose sensitive data and permit unauthorized access. Addressing these vulnerabilities is crucial for maintaining the security and functionality of the services they support. 
\end{tcolorbox}

\subsection{\textbf{DevOps and CI/CD Practices}}
In microservice architectures, DevOps and CI/CD refer to the processes and practices that integrate development and operations teams to improve the efficiency, security, and agility of software delivery \cite{throner2021advanced}. This includes setting up pipelines for continuous integration (CI) and continuous deployment (CD), managing Infrastructure as Code (IaC), and implementing protocols for managing secrets and credentials. Security in this area is essential to automate and secure the delivery pipeline and to dynamically investigate security throughout the development lifecycle \cite{CarneiroSchmelmer2016Microservices}.

\subsubsection{CI/CD Pipeline}

The CI/CD pipeline is a critical component of microservice architectures that automates the integration and deployment processes. It involves building, testing, and deploying code changes continuously and efficiently. Securing the CI/CD pipeline is essential, as vulnerabilities in this process lead to unauthorized access, data breaches, and service disruptions, thereby compromising the integrity and functionality of the microservices \cite{SurveyMSAAdvancedManufacturing}.

\textit{V$_{107}$ - Vulnerable CI/CD Tools:} This vulnerability arises from integration tools within microservice architectures that lack essential security controls such as authentication, authorization, encryption, and monitoring \cite{gotz2018challenges}. These shortcomings make the tools exploitable, leading to unauthorized access, data breaches, and system compromise. For example, CVE-2022-28145 highlights the risks associated with insecure integration tools \cite{mitre0107}.

\vspace{0.03in}

\textit{V$_{108}$ - CI/CD Misconfigurations:} This vulnerability occurs due to misconfigurations in CI/CD pipelines within microservice architectures, which are essential for automating build, test, and deployment processes \cite{Balalaie2016MicroservicesDevOps}. When these pipelines lack adequate security measures such as access controls, secrets management, and vulnerability scanning, they become vulnerable to exploitation. This leads to unauthorized access, data breaches, and service disruptions. For instance, CVE-2022-22984 shows the dangers of CI/CD misconfigurations \cite{mitre0108}.

\vspace{0.03in}

\textit{V$_{109}$ - Lack of Pipeline Security Controls:} This vulnerability arises within microservice architectures due to the absence of security controls in development pipelines. These pipelines automate crucial software development processes and, without essential security measures like access controls and vulnerability scanning, are vulnerable to exploitation \cite{SurveyMSAAdvancedManufacturing}. This leads to unauthorized access, data breaches, and system compromise. For example, CVE-2019-16574 shows the consequences of the absence of security controls for CI/CD pipelines \cite{mitre0109}.

\vspace{0.03in}

\textit{V$_{110}$ - Deploying Vulnerable Code:} This vulnerability occurs within microservice architectures when vulnerable or untested code is deployed into production. Inadequate security controls or validation leads to exploitable vulnerabilities, risking data breaches, service disruptions, and unauthorized access. Exploitation of this vulnerability compromises an entire microservice infrastructure \cite{MazzaraOberSalaun2018STAF}. For example, CVE-2020-5252 underlines the impact of insecure coding practices in DevOps environments \cite{mitre0110}.

\begin{tcolorbox}[left=0pt, top=0pt, right=0pt, bottom=0pt, boxrule=0.75pt]
Securing DevOps and CI/CD processes is critical in microservice architectures to prevent vulnerabilities such as insecure CI/CD tools, pipeline misconfigurations, lack of security controls, and deploying vulnerable code. These issues lead to unauthorized access, data breaches, and system compromises. Effective security measures and vigilant configuration management are essential to safeguard the integrity and functionality of automated development pipelines.
\end{tcolorbox}

\subsubsection{Infrastructure as Code}

Infrastructure as Code (IaC) in microservices involves managing and provisioning computing infrastructure through machine-readable scripts rather than manual processes. This practice ensures consistency and efficiency in deployment. However, improper securing of IaC scripts and configurations leads to significant security risks \cite{war2023security}. Ensuring robust security measures and proper configuration management is essential to protect the integrity and functionality of the infrastructure \cite{SurveyMSAAdvancedManufacturing}.

\textit{V$_{111}$ - Insecure IaC Scripts:} This vulnerability arises from weaknesses or misconfigurations in IaC scripts within microservice architectures \cite{war2023security, NkomoCoetzee2019SecureMicroservices}. Issues such as improper permission settings or insecure configurations lead to significant risks like data exposure, unauthorized access, or entire environment compromise. For example, CVE-2024-38346 shows the impact of insecure IaC scripts in DevOps practices \cite{mitre0111}.

\vspace{0.03in}

\textit{V$_{112}$ - Misconfigured Automated Deployment:} This vulnerability occurs when automated deployment processes result in insecure or improperly configured environments within microservice architectures \cite{RademacherSorgallaSachweh2018}. Weak authentication, exposed data, or insecure access controls severely impact data security, service availability, or system integrity. For example, CVE-2022-23635 shows the broader impact of misconfigured deployment of microservice infrastructures \cite{mitre0112}.

\vspace{0.03in}

\textit{V$_{113}$ - Hardcoded Secrets in IaC:} This vulnerability arises when sensitive information is embedded directly in IaC scripts without proper security measures \cite{SurveyMSAAdvancedManufacturing}. Storing passwords, credentials, or keys in scripts without encryption leads to unauthorized access or data breaches. A potential example is CVE-2024-33891, which emphasizes the consequences of hardcoded secrets in IaC \cite{mitre0113}.

\vspace{0.03in}

\textit{V$_{114}$ - No IaC Version Control:} This vulnerability occurs when IaC scripts lack version control, leading to potential configuration drift or deployment inconsistencies \cite{SurveyMSAAdvancedManufacturing}. The inability to manage, collaborate on, or audit infrastructure changes severely affects the maintainability and security of infrastructure configurations. A potential example is CVE-2024-41110, which outlines the risks of flaws in different versions of Docker used for software containerization \cite{mitre0114}.

\begin{tcolorbox}[left=0pt, top=0pt, right=0pt, bottom=0pt, boxrule=0.75pt]
Securing  IaC is vital in microservice architectures to prevent vulnerabilities like insecure IaC scripts, misconfigured automated deployments, hardcoded secrets, and lack of version control. These issues lead to unauthorized access, data breaches, and compromised environments. Implementing robust security measures and proper configuration management is essential to protect the integrity and functionality of IaC, as evidenced by various CVEs highlighting the severe impact of these vulnerabilities.

\end{tcolorbox}

\subsubsection{Secrets Management}

Secrets Management in microservices focuses on handling sensitive information such as passwords and security tokens securely. It involves storing, managing, and rotating secrets to prevent unauthorized access and data breaches. Proper secrets management is crucial to maintaining the security of microservices, as poor practices lead to significant security breaches and unauthorized access \cite{VelepuchaFlores2023Microservices}.

\textit{V$_{115}$ - Poor Secrets Rotation:} This vulnerability arises from inadequate management of sensitive information like passwords and API tokens in microservice architectures \cite{Billawa2022MicroserviceSecurity}. Poor storage mechanisms and rotation policies, such as storing secrets in plaintext and neglecting regular rotation, lead to unauthorized access, data breaches, and system compromise. For example, CVE-2024-41129 illustrates the broader impact of inadequate management of secrets \cite{mitre0115}.

\vspace{0.03in}

\textit{V$_{116}$ - Lack of Centralized Secrets Management:} This vulnerability occurs due to the absence of a centralized system for managing sensitive information in microservice architectures \cite{Billawa2022MicroserviceSecurity}. Ad-hoc management practices lacking protection or control severely increase the risk of exposure and misuse of secrets across the environment. For example, CVE-2023-30617 highlights the dangers of inadequate secrets management \cite{mitre0116}.

\vspace{0.03in}

\textit{V$_{117}$ - Secrets in Version Control:} This vulnerability arises when sensitive information is stored in version control repositories within microservice architectures \cite{Belair2019ContainerSecurity}. Storing secrets in code or configuration files committed to VCS leads to unauthorized access, data breaches, and significant security breaches. A potential example is CVE-2024-41122, which underscores the risks of poor secret management practices \cite{mitre0117}.

\vspace{0.03in}

\textit{V$_{118}$ - Unmonitored Secrets Access:} This vulnerability occurs when access to sensitive information is not tracked via logs in microservice architectures \cite{Billawa2022MicroserviceSecurity}. Insufficient monitoring increases the risk of undetected breaches or unauthorized activities, leading to data breaches and unauthorized system access. An example is CVE-2018-3831, which highlights the importance of logging and monitoring secrets \cite{mitre0118}.

\begin{tcolorbox}[left=0pt, top=0pt, right=0pt, bottom=0pt, boxrule=0.75pt]
Effective secrets management is crucial in microservice architectures to prevent vulnerabilities like poor secrets rotation, lack of centralized secrets management, storing secrets in version control, and unmonitored secrets. These issues lead to unauthorized access and significant data breaches. Implementing robust storage mechanisms, regular rotation policies, centralized management, and comprehensive monitoring is essential to safeguard sensitive information and maintain system security. 
\end{tcolorbox}

\subsubsection{ Testing and Analysis}

Testing and analysis encompass conducting thorough security evaluations and assessments throughout the entire lifecycle of developing and deploying microservice architectures \cite{nkomo2019software}. This includes both static and dynamic testing to identify potential security gaps. Proper testing practices are essential to detect and address vulnerabilities early, ensuring the security and integrity of microservice architectures \cite{Soylemez2022FeatureDriven}.

\textit{V$_{119}$ - Lack of Automated Testing:} This vulnerability arises from inadequate automated test coverage during DevOps operations in microservice architectures \cite{prabath2020}. Without comprehensive automated testing, software defects, vulnerabilities, or regressions may go undetected, increasing the risk of deploying insecure services and resulting in service disruptions or data breaches. A potential example is CVE-2021-26084, which unauthenticated attackers to execute attacks due to a lack of automated testing \cite{mitre0119}.

\vspace{0.03in}

\textit{V$_{120}$ - Poor Manual Security Reviews:} This vulnerability occurs when there are inadequate manual security reviews in DevOps operations within microservice architectures \cite{newman2015}. Neglecting manual reviews leads to overlooked security risks, increasing the likelihood of undetected vulnerabilities or design flaws. A potential example is CVE-2024-5687, which highlights the importance of thorough security checks \cite{mitre0120}.

\vspace{0.03in}

\textit{V$_{121}$ - Ignoring Test Results:} This vulnerability arises when issues identified in testing phases are ignored or delayed in response within microservice architectures \cite{newman2015}. Not addressing test findings critically endangers microservice architectures, leading to the deployment of flawed services and resulting in service disruptions or security breaches. A potential example is CVE-2021-40539, which underlines the consequences of ignoring test results or delaying response to security issues \cite{mitre0121}.

\vspace{0.03in}

\textit{V$_{122}$ - Outdated Testing Tools:} This vulnerability occurs due to relying on outdated testing tools that lack recent updates or have known vulnerabilities within microservice architectures \cite{prabath2020}. Using tools that are not up-to-date leads to undetected issues. A potential example is CVE-2024-3234, which showcases the risks associated with using outdated components \cite{mitre0122}.

\begin{tcolorbox}[left=0pt, top=0pt, right=0pt, bottom=0pt, boxrule=0.75pt]
Robust security testing and analysis are indispensable in microservice architectures to uncover and address potential vulnerabilities. Key issues such as lack of automated testing, insufficient manual security reviews, ignoring test results, and using outdated testing tools leave microservices exposed to threats, leading to service disruptions and data breaches. Implementing comprehensive, up-to-date security practices is essential to maintain system integrity and ensure reliable, secure microservice operations.\end{tcolorbox}

\begin{figure*}[t!]
    \centering
    \includegraphics[width=0.65\paperwidth]{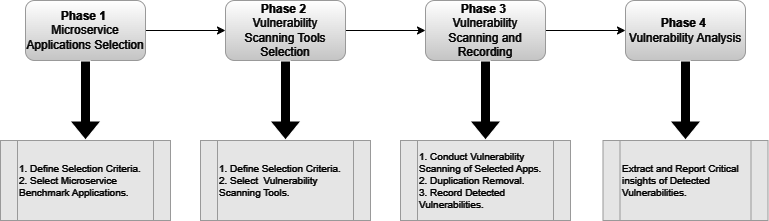}
    \caption{Empirical Evaluation Process.}
    \label{fig: eval_proc}
\end{figure*}

\subsubsection{\textit{DevSecOps Integration}}
DevSecOps Integration in microservices involves implementing security practices within Development Operations (DevOps) to address vulnerabilities that may arise during the development lifecycle \cite{chandramouli2022implementation}. It emphasizes the importance of integrating security measures throughout the development process to mitigate risks and ensure robust protection of microservice architectures. This integration helps in identifying and addressing security issues early, thereby reducing the potential for security breaches and enhancing the overall security posture of the microservices \cite{SurveyMSAAdvancedManufacturing}.

\textit{V$_{123}$ - Inadequate Security Training:} This vulnerability occurs when there is an absence or inadequacy of training programs or resources for DevSecOps teams working with microservice architectures. Without proper training initiatives, team members may lack the necessary skills, knowledge, and awareness to effectively develop, deploy, and maintain secure microservice-based systems \cite{MateusCoelho2021SecuringMicroservices}. This increases the risk of security vulnerabilities, performance issues, or compliance violations. A potential example is CVE-2021-34523, which underscores the importance of continuous security training and awareness among DevSecOps teams to prevent the late discovery of critical issues \cite{mitre0123}.

\vspace{0.03in}

\textit{V$_{124}$ - Neglecting Early Security:} This vulnerability arises when security considerations are not integrated within the early stages of the software development lifecycle in microservice architectures \cite{MateusCoelho2021SecuringMicroservices}. Neglecting or delaying security requirements, threats, or risks during planning, design, or requirements-gathering phases introduces security vulnerabilities, design flaws, or compliance issues. A potential example is CVE-2003-0550, which highlights the importance of security by design approach \cite{mitre0124}.

\vspace{0.03in}

\textit{V$_{125}$ - Inconsistent Security Practices:} This vulnerability occurs due to the irregular or uneven implementation of security measures within microservice architectures. Lack of standardized security practices across different components, services, and teams creates security gaps, misconfigurations, or vulnerabilities \cite{MateusCoelho2021SecuringMicroservices}. A potential example is CVE-2018-13379, which occurred due to improper security practices \cite{mitre0125}.

\vspace{0.03in}

\textit{V$_{126}$ - Poor Security-DevOps Collaboration:} This vulnerability arises from inadequate cooperation and communication between security teams and DevOps teams within microservice architectures \cite{koskinen2019devsecops}. A lack of integration between security practices and DevOps processes results in security considerations being overlooked during the development, deployment, and operation of services \cite{MateusCoelho2021SecuringMicroservices}.

\begin{tcolorbox}[left=0pt, top=0pt, right=0pt, bottom=0pt, boxrule=0.75pt]
DevSecOps integration is vital for securing microservice architectures. Key vulnerabilities such as inadequate security training, neglecting early security integration, inconsistent security practices, and poor collaboration between security and DevOps teams lead to significant risks. Addressing these issues through comprehensive training, early inclusion of security measures, standardized practices, and improved teamwork is essential to mitigate vulnerabilities and ensure the robustness of microservices. 
\end{tcolorbox}

\begin{table*}[t!]
    \centering
    \caption{Benchmark Microservice Applications.}
    \renewcommand{\arraystretch}{1}
    \begin{tabular}{lp{13cm}} 
        \hline
        \textbf{Benchmark Microservice Application} & \textbf{Description} \\ \hline
        Book Management System \cite{wolff2023microservice} & A comprehensive book management system developed using Java and HTML. This application demonstrates standard CRUD operations and integrates various microservices to manage books, authors, and user interactions. \\ \hline
        Cloud Native E-Shop \cite{ravuri2023ecommerce} & A robust cloud-native e-commerce platform built using TypeScript, JavaScript, Java, Python, HCL, and HTML. It showcases advanced features such as user authentication, product catalog management, and order processing. \\ \hline
        Spring Boot Application \cite{in28minutes2023springMicroservices} & A suite of web services developed using Spring Boot in Java, JavaScript, and HTML. These applications illustrate the use of microservices for building scalable and maintainable web services. \\ \hline
        Knowledge Base Application \cite{cscdevops2023app} & A streamlined application developed with JavaScript (Node.js and Express.js) and MongoDB. It focuses on efficient data storage and retrieval, providing a simple yet effective solution for managing knowledge articles. \\ \hline
    \end{tabular}
    \label{table:micro}
\end{table*}


\section{Empirical Analysis of Microservice Vulnerabilities}\label{sec5}

In this section, we present an in-depth empirical analysis of security vulnerabilities within microservice architectures. This analysis is conducted through a systematic and cost-effective approach, selecting benchmark microservice applications and suitable vulnerability scanning tools to ensure the reliability and reproducibility of our findings. Figure \ref{fig: eval_proc} depicts our empirical evaluation process, which consists of four distinct phases explained as follows.

\subsection{Phase 1: Microservice Applications Selection }
To maintain cost-effectiveness and ensure accessibility, we select open-source microservice benchmark applications. This approach minimizes expenses and facilitates result verification by other researchers, thereby enhancing the credibility of our findings. We emphasize the diversity and currency of the chosen applications; diverse applications help mitigate bias in vulnerability scanning, while up-to-date codebases ensure accurate identification of relevant vulnerabilities. Table \ref{table:micro} presents an overview of the benchmark microservice applications selected for this study, chosen for their diverse functionality, technology stack, and architecture to ensure comprehensive vulnerability coverage. The Book Management System (Java, HTML) showcases CRUD operations with integrated microservices. The Cloud Native E-Shop (TypeScript, JavaScript, Java, Python, HCL, HTML) includes user authentication, product catalog management, and order processing. The Spring Boot Application suite (Java, JavaScript, HTML) utilizes the Spring Microservices framework for scalable web services. Finally, the Knowledge Base Application (Node.js, Express.js, MongoDB) emphasizes efficient data storage and retrieval. This diversity mitigates bias and ensures the relevance and applicability of our findings across various real-world scenarios.

\subsection{Phase 2: Vulnerability Scanning Tools Selection}
We establish the specific criteria to guide the selection of vulnerability scanning tools, ensuring a robust and unbiased evaluation:

\begin{itemize}[leftmargin=*]
    \item \textbf{Open-Source Availability.} Tools have to be free and accessible to align with budget constraints.
    \item \textbf{Community Recognition.} Widely recognized tools in the research community to ensure credible and verifiable results.
    \item \textbf{Comprehensive Database.} Tools with extensive vulnerability databases are preferred for thorough scanning and identification of known vulnerabilities.
    \item \textbf{User-Friendliness.} Tools need to be user-friendly to streamline the scanning process, allowing us to focus on assessing the tools' accuracy rather than usability.
\end{itemize}

Initially, we selected Deepsource, SNYK, and Trivy. However, Deepsource is found unsuitable as it provides Common Weakness Enumeration (CWE) numbers instead of Common Vulnerabilities and Exposures (CVE) numbers, creating inconsistencies. Consequently, other tools such as OSV Scanner, Semgrep, OWASP ZAP, and OpenVas are tested but exhibit significant shortcomings. Ultimately, we select OWASP Dependency Check alongside Trivy and SNYK.

\subsection{Phase 3: Vulnerability Scanning and Recording}
Upon finalizing the benchmark applications and scanning tools, we conduct comprehensive vulnerability scans using Trivy, SNYK, and OWASP Dependency Check. For each microservice benchmark application, we meticulously record the detected vulnerabilities. This vulnerability data collection is managed in Microsoft Excel, with separate spreadsheets created for each application and each scanning tool. This organization provides a clear and structured overview of the results, facilitating easy analysis and comparison of the findings. Then, we develop an algorithm to identify unique vulnerabilities from the recorded data. This algorithm segregates vulnerabilities by severity, as defined by the Common Vulnerability Scoring System (CVSS), into four categories: CRITICAL, HIGH, MEDIUM, and LOW. The unique vulnerabilities and their frequencies are compiled into a two-column table and used to update the Excel sheets, adding a severity column for comprehensive analysis. The cleaned and analyzed data form the dataset for further research. Detailed information is available in the repository of this study \cite{repository2024}. Table \ref{tab:final_vulnerabilities} presents the number of unique vulnerabilities identified for each application by the three scanning tools. A total of 1667 vulnerabilities are discovered across all four applications, distributed as follows: 618 in the Book Management System, 128 in the Cloud Native E-Shop, 477 in the Spring Boot application, and 444 in the Knowledge Base application.

\begin{table}[t!]
\centering
\caption{Number of Vulnerabilities Identified}
\label{tab:final_vulnerabilities}
\begin{tabular}{p{2.8cm}p{0.7cm}p{0.7cm}p{1.5cm}p{0.7cm}} 

\toprule
\textbf{Application} & \textbf{SNYK} & \textbf{Trivy} & \textbf{OWASP DC}  & \textbf{Total} \\
\midrule
Book Management App & 545 & 49 & 24  & 618 \\
Cloud Native E-Shop & 96 & 17 & 15  & 128 \\
Spring Boot & 208 & 127 & 142  & 477 \\
Knowledge Base & 390 & 21 & 33  & 444 \\
\bottomrule
\end{tabular}
\end{table}

\subsection{Phase 4: Vulnerability Analysis}

Our empirical analysis provides critical insights into the security vulnerabilities in microservice architectures. These insights underscore the importance of a comprehensive security strategy tailored to the specific characteristics of microservices.

\noindent \textbf{Diversity in Vulnerability Detection.} Our analysis highlights significant differences in the vulnerabilities identified by various scanning tools, underscoring the importance of using multiple tools to ensure comprehensive security coverage. Each tool has unique capabilities, allowing for a broader detection range. For instance, Trivy and OWASP Dependency Check detect different vulnerabilities, illustrating the limitations of relying on a single tool. This emphasizes the necessity of a multifaceted approach to vulnerability scanning in complex microservice environments.

\noindent \textbf{Tool Effectiveness.} Comparing the findings from different tools shows significant differences in their effectiveness in various contexts. For instance, SNYK is particularly effective in identifying vulnerabilities related to outdated dependencies, while OWASP Dependency Check excels at finding configuration-related issues. This comparison highlights the importance of selecting tools based on the specific needs and characteristics of the microservices being analyzed.

\noindent \textbf{Role of Automated Tools.} Automated vulnerability scanning tools are crucial for identifying potential security issues. However, manual reviews are necessary to contextualize and accurately interpret the results. Combining automated tools with manual efforts provides a more comprehensive security assessment, especially in complex microservice environments.

\noindent \textbf{Severity Distribution.} Understanding the distribution of vulnerability severity is essential for prioritizing remediation efforts. Our analysis categorizes vulnerabilities into CRITICAL, HIGH, MEDIUM, and LOW levels, helping prioritize resources effectively. Addressing CRITICAL vulnerabilities is crucial as they pose immediate risks, potentially leading to severe breaches. In contrast, LOW severity vulnerabilities can be addressed in a more phased manner, optimizing resource allocation.

\noindent \textbf{Continuous Monitoring.} The dynamic nature of microservices requires continuous monitoring and regular vulnerability assessments. Static security measures are inadequate in such a rapidly changing environment. Integrating vulnerability scanning into the continuous integration and continuous deployment pipeline is vital. This integration ensures that new code deployments do not introduce new vulnerabilities and that existing vulnerabilities are regularly reassessed.

\noindent \textbf{Impact of Open-Source Components.} The extensive use of open-source components in microservice architectures presents significant security challenges. Many identified vulnerabilities are associated with these components, highlighting the need for regular updates and patches. While open-source software offers benefits like cost efficiency and community support, it requires diligent management to ensure security.

\noindent \textbf{Importance of Secure Development Practices.} The analysis stresses the need for secure development practices throughout the software lifecycle. This includes comprehensive code reviews, secure coding standards, and developer training. Incorporating security from the beginning helps minimize the introduction of vulnerabilities, thereby enhancing the overall security of the system.

\noindent \textbf{Community and Vendor Support.} Leveraging community and vendor support enhances vulnerability management. Community involvement ensures timely updates and best practices, while vendor support provides additional resources and expertise, especially for complex security challenges.

\noindent \textbf{False Positives and Negatives.} Instances of false positives and negatives underscore the limitations of automated tools. It is essential to validate the results with manual inspection to ensure accurate vulnerability assessments.

\noindent \textbf{Impact on Different Microservices.} The impact of vulnerabilities varies significantly among different microservices. For example, the BMS Microservice had the highest number of vulnerabilities, possibly due to its complex architecture and extensive use of third-party libraries. Understanding the unique characteristics of each microservice is crucial for tailoring security measures to address specific risks.

\noindent \textbf{Resource Allocation for Remediation.} Effective vulnerability management requires prioritizing remediation efforts based on severity and potential impact. Our analysis supports a risk-based approach, focusing on the most critical vulnerabilities first, which helps in the efficient use of resources and enhances overall security.

\begin{tcolorbox}[left=0pt, top=0pt, right=0pt, bottom=0pt, boxrule=0.75pt]
In conclusion, our analysis of empirically identified vulnerabilities highlights the importance of a comprehensive and proactive approach to vulnerability management in microservice architectures. By integrating diverse tools, continuous monitoring, and a culture of security awareness, organizations can effectively protect their microservice architectures against evolving threats. This approach not only addresses current vulnerabilities but also strengthens the resilience of microservice-based applications against future challenges. 
\end{tcolorbox}

\section{Discussion}\label{sec6}

In this section, we provide a detailed analysis of our research findings by examining the proposed vulnerability taxonomy (Section \ref{sec4}) in relation to the empirical analysis of microservice security (Section \ref{sec5}). We map the vulnerabilities categorized in our taxonomy against those identified through the empirical analysis of different microservice applications. This alignment not only serves to validate the proposed taxonomy but also provides a more comprehensive understanding of security issues within microservice architectures. The mapping process involves a thorough comparison between the theoretical vulnerabilities outlined in the taxonomy and the practical vulnerabilities observed in real-world microservice systems. By doing so, we assess the accuracy and relevance of the taxonomy in reflecting real-world security challenges faced in microservice environments. This validation is crucial for ensuring that the taxonomy is grounded in practical experiences and is useful for real-world applications. In the following sections, we detail the mapping process, analyze its findings, and provide insights for both researchers and practitioners.

\subsection{Mapping Process}

We map the proposed taxonomy vulnerabilities to those empirically identified through a structured three-step approach. \textit{Step 1: Manual Comparison.} In the initial step, we manually compare each vulnerability detected by the scanning tools with the vulnerabilities listed in the proposed taxonomy. This involves a detailed review of each empirical vulnerability to determine the closest match within the taxonomy framework. We examine the characteristics, causes, and impacts of each vulnerability through the NVD database to find the most relevant taxonomy category and vulnerability. \textit{Step 2: Literature Review.} When a direct match is not immediately apparent, we consult existing literature to identify similar vulnerabilities that may have been categorized differently or discussed in other contexts. This helps to align the empirical data with the taxonomy by referencing established knowledge and categorizations.
\textit{Step 3: Validation and Consensus.} In the final step, we validate the mapping through discussions and consensus-building among our research team. This collaborative process ensures that the mapping is both consistent and comprehensive. This iterative process refines the taxonomy to better reflect real-world findings. This three-step approach ensures a thorough and accurate alignment of vulnerabilities, bridging theoretical taxonomy with practical empirical data, and enhancing the applicability and relevance of our proposed taxonomy for microservice vulnerabilities.

As detailed in Section \ref{sec5}, a total of 1,667 vulnerabilities were identified across four applications using three scanning tools. These were segmented into 618 vulnerabilities in the Book Manager System, 128 in the Cloud-native E-Shop Application, 477 in the Spring Boot Application, and 444 in the Knowledge Base Application. To facilitate the mapping, we created four separate Excel sheets within a single document, each corresponding to one of the benchmark applications. Each sheet contains proposed vulnerabilities from the taxonomy, CVE numbers of identified vulnerabilities through each of the three scanning tools, and short descriptions of each identified vulnerability. For more details, please refer to our database available at \cite{repository2024}. Each CVE number is mapped to one of the 126 proposed vulnerabilities in the taxonomy. For example, CVE-2021-39148, an injection vulnerability, is mapped to our taxonomy vulnerability V$_{99}$, which covers injection vulnerab-

\onecolumn
{\fontsize{4.5pt}{5pt}\selectfont
\begin{longtable}{|l|c|c|c|c|c|c|c|c|c|c|c|c|c|c|c|c|}
\caption{Mapping of Proposed Taxonomy to Empirically Identified Vulnerabilities.} \label{tab: mapping}\\
\hline
\multirow{2}{*}{\textbf{\normalsize Taxonomy Vulnerabilities}} & \multicolumn{3}{c|}{\scriptsize Book Mgt System} & \multicolumn{3}{c|}{\scriptsize Cloud Native Eshop} & \multicolumn{3}{c|}{\scriptsize Spring Boot} & \multicolumn{3}{c|}{\scriptsize Knowledge Base} & \multicolumn{3}{c|}{\scriptsize {Tools}} & \scriptsize Identified \\ \cline{2-16}
 & SNYK & Trivy & OWASP & SNYK & Trivy & OWASP & SNYK & Trivy & OWASP & SNYK & Trivy & OWASP & SNYK & Trivy & OWASP & \scriptsize Vulns. \\ 
 \hline
\endfirsthead
\hline
V$_{1}$: Exposed API Endpoints. & 4 & 0 & 0 & 1 & 0 & 0 & 2 & 0 & 0 & 0 & 0 & 1 & 7 & 0 & 1 & 8 \\
V$_{2}$: Accidental Exposure of Sensitive API Endpoints. & 6 & 0 & 0 & 0 & 0 & 0 & 1 & 0 & 0 & 0 & 0 & 0 & 7 & 0 & 0 & 7 \\
V$_{3}$: Untrusted Third-Party APIs. & 0 & 0 & 0 & 0 & 0 & 0 & 0 & 0 & 1 & 0 & 0 & 0 & 0 & 0 & 1 & 1 \\
V$_{4}$: Weak API Authentication. & 5 & 0 & 0 & 2 & 1 & 0 & 2 & 0 & 0 & 0 & 0 & 1 & 9 & 1 & 1 & 11 \\
V$_{5}$: Insecure API Deserialization. & 0 & 0 & 0 & 0 & 0 & 0 & 0 & 0 & 1 & 0 & 0 & 1 & 0 & 0 & 2 & 2 \\
V$_{6}$: Misconfigured API Gateways. & 4 & 0 & 0 & 1 & 0 & 0 & 0 & 0 & 0 & 0 & 0 & 1 & 5 & 0 & 1 & 6 \\
V$_{7}$: Service Registration Tampering. & 3 & 0 & 0 & 1 & 0 & 0 & 2 & 1 & 0 & 0 & 0 & 0 & 6 & 1 & 0 & 7 \\
V$_{8}$: Unauthorized Service Discovery Access. & 2 & 0 & 0 & 1 & 0 & 0 & 2 & 1 & 0 & 0 & 0 & 0 & 5 & 1 & 0 & 6 \\
V$_{9}$: Service Registration Validation Gaps. & 0 & 0 & 0 & 0 & 0 & 0 & 0 & 0 & 2 & 3 & 0 & 0 & 3 & 0 & 2 & 5 \\
V$_{10}$: Unauthorized Service Deregistration. & 2 & 0 & 0 & 1 & 0 & 0 & 3 & 1 & 0 & 0 & 0 & 0 & 6 & 1 & 0 & 7 \\
V$_{11}$: Replay Attacks on Service Requests. & 3 & 0 & 0 & 0 & 0 & 0 & 0 & 0 & 0 & 0 & 0 & 0 & 3 & 0 & 0 & 3 \\
V$_{12}$: Service Impersonation. & 0 & 0 & 0 & 0 & 0 & 0 & 0 & 0 & 1 & 0 & 0 & 0 & 0 & 0 & 1 & 1 \\
V$_{13}$: Poor Network Segmentation. & 0 & 0 & 0 & 0 & 0 & 0 & 1 & 0 & 1 & 0 & 0 & 0 & 1 & 0 & 1 & 2 \\
V$_{14}$: Misconfigured Service Mesh. & 0 & 0 & 0 & 0 & 0 & 0 & 1 & 0 & 2 & 0 & 0 & 0 & 1 & 0 & 2 & 3 \\
V$_{15}$: Misconfigured Network Access. & 1 & 0 & 0 & 0 & 0 & 0 & 0 & 0 & 0 & 0 & 0 & 0 & 1 & 0 & 0 & 1 \\
V$_{16}$: Improper Firewall Configuration. & 0 & 0 & 0 & 0 & 0 & 0 & 0 & 0 & 2 & 0 & 0 & 0 & 0 & 0 & 2 & 2 \\
V$_{17}$: Segmentation Bypass. & 0 & 0 & 0 & 0 & 0 & 0 & 0 & 0 & 2 & 0 & 0 & 0 & 0 & 0 & 2 & 2 \\
V$_{18}$: Default Network Configurations. & 4 & 0 & 0 & 0 & 0 & 0 & 0 & 0 & 0 & 0 & 0 & 0 & 4 & 0 & 0 & 4 \\
V$_{19}$: Weak In-Transit Data Encryption. & 0 & 0 & 0 & 0 & 0 & 0 & 0 & 0 & 2 & 0 & 0 & 0 & 0 & 0 & 2 & 2 \\
V$_{20}$: Faulty Certificate Validation. & 0 & 0 & 0 & 0 & 0 & 0 & 0 & 0 & 1 & 0 & 0 & 0 & 0 & 0 & 1 & 1 \\
V$_{21}$: In-Transit Metadata Exposure. & 6 & 0 & 0 & 0 & 0 & 0 & 0 & 0 & 0 & 0 & 0 & 0 & 6 & 0 & 0 & 6 \\
V$_{22}$: Encryption Key Mismanagement. & 5 & 0 & 0 & 0 & 0 & 0 & 0 & 0 & 0 & 0 & 0 & 0 & 5 & 0 & 0 & 5 \\
V$_{23}$: Inadequate Network Certificate Validation. & 1 & 0 & 0 & 0 & 0 & 0 & 0 & 0 & 0 & 0 & 0 & 0 & 1 & 0 & 0 & 1 \\
V$_{24}$: Hardcoded Encryption Keys. & 8 & 0 & 0 & 0 & 0 & 0 & 0 & 0 & 0 & 0 & 0 & 0 & 8 & 0 & 0 & 8 \\
V$_{25}$: Absence of Dynamic Rate Limiting. & 11 & 1 & 1 & 2 & 0 & 0 & 3 & 0 & 2 & 3 & 0 & 0 & 19 & 1 & 3 & 23 \\
V$_{26}$: Lack of IP-Based Rate Limiting. & 3 & 0 & 0 & 2 & 0 & 0 & 1 & 2 & 2 & 6 & 0 & 0 & 12 & 2 & 2 & 16 \\
V$_{27}$: Generic Rate Limiting Policies. & 7 & 2 & 0 & 2 & 0 & 0 & 1 & 4 & 4 & 9 & 0 & 1 & 19 & 6 & 5 & 30 \\
V$_{28}$: Misconfigured Rate Limits. & 5 & 1 & 0 & 3 & 3 & 1 & 1 & 0 & 2 & 12 & 0 & 1 & 21 & 4 & 4 & 29 \\
V$_{29}$: Token Bucket Overflow. & 4 & 0 & 0 & 3 & 0 & 0 & 1 & 0 & 4 & 6 & 0 & 1 & 14 & 0 & 5 & 19 \\
V$_{30}$: Weak Database Encryption. & 9 & 5 & 0 & 3 & 1 & 0 & 3 & 2 & 1 & 9 & 1 & 0 & 24 & 9 & 1 & 34 \\
V$_{31}$: Insufficient Database Hardening. & 10 & 3 & 1 & 3 & 1 & 0 & 10 & 2 & 1 & 9 & 2 & 0 & 32 & 8 & 2 & 42 \\
V$_{32}$: Exposure of Default Database Credentials. & 15 & 6 & 3 & 5 & 2 & 1 & 17 & 7 & 6 & 18 & 1 & 2 & 55 & 16 & 12 & 83 \\
V$_{33}$: Sensitive Data in Error Messages. & 6 & 0 & 0 & 0 & 0 & 0 & 4 & 0 & 0 & 2 & 0 & 0 & 12 & 0 & 0 & 12 \\
V$_{34}$: Absence of Data Integrity Checks. & 2 & 2 & 1 & 1 & 0 & 0 & 5 & 0 & 1 & 10 & 0 & 1 & 18 & 2 & 3 & 23 \\
V$_{35}$: SQL Injection. & 5 & 0 & 0 & 5 & 1 & 0 & 0 & 1 & 0 & 16 & 0 & 1 & 26 & 2 & 1 & 29 \\
V$_{36}$: Cross-Site Scripting. & 57 & 5 & 2 & 0 & 0 & 0 & 21 & 15 & 6 & 1 & 0 & 2 & 79 & 20 & 10 & 109 \\
V$_{37}$: Command Injection. & 33 & 2 & 2 & 5 & 0 & 3 & 14 & 8 & 10 & 19 & 3 & 1 & 71 & 13 & 16 & 100 \\
V$_{38}$: Insecure Data Deserialization. & 3 & 0 & 0 & 0 & 0 & 0 & 0 & 1 & 0 & 0 & 0 & 0 & 3 & 1 & 0 & 4 \\
V$_{39}$: Direct Object Reference. & 16 & 2 & 1 & 1 & 0 & 1 & 1 & 0 & 3 & 7 & 1 & 3 & 25 & 3 & 8 & 36 \\
V$_{40}$: Excessive Privilege Granting. & 13 & 1 & 1 & 1 & 0 & 0 & 4 & 2 & 0 & 0 & 0 & 0 & 18 & 3 & 1 & 22 \\
V$_{41}$: Hardcoded Credentials Exposure. & 1 & 0 & 0 & 2 & 0 & 0 & 0 & 0 & 0 & 0 & 0 & 0 & 3 & 0 & 0 & 3 \\
V$_{42}$: Privilege Escalation. & 11 & 1 & 0 & 0 & 0 & 0 & 1 & 2 & 0 & 0 & 0 & 0 & 12 & 3 & 0 & 15 \\
V$_{43}$: Race Condition Exploitation. & 1 & 0 & 0 & 1 & 0 & 0 & 0 & 0 & 0 & 0 & 0 & 1 & 2 & 0 & 1 & 3 \\
V$_{44}$: Inadequate Data Transaction Management. & 1 & 0 & 0 & 2 & 1 & 0 & 1 & 0 & 0 & 0 & 0 & 2 & 4 & 1 & 2 & 7 \\
V$_{45}$: Weak Data Synchronization. & 0 & 0 & 0 & 0 & 0 & 1 & 4 & 1 & 0 & 6 & 0 & 2 & 10 & 1 & 3 & 14 \\
V$_{46}$: Concurrent Data Access Mismanagement. & 3 & 0 & 0 & 0 & 0 & 1 & 0 & 1 & 1 & 4 & 1 & 0 & 7 & 2 & 2 & 11 \\
V$_{47}$: Irregular Backups Maintenance. & 5 & 0 & 0 & 0 & 0 & 0 & 4 & 3 & 0 & 4 & 0 & 0 & 13 & 3 & 0 & 16 \\
V$_{48}$: Insecure Backup Storage. & 2 & 0 & 0 & 0 & 0 & 0 & 1 & 0 & 0 & 7 & 1 & 0 & 10 & 1 & 0 & 11 \\
V$_{49}$: Lack of Backup Validation. & 1 & 0 & 0 & 1 & 0 & 1 & 3 & 2 & 0 & 4 & 1 & 0 & 9 & 3 & 1 & 13 \\
V$_{50}$: Improper Backup Disposal. & 1 & 0 & 0 & 0 & 0 & 0 & 0 & 1 & 1 & 1 & 0 & 0 & 2 & 1 & 1 & 4 \\
V$_{51}$: Weak Authentication Protocols. & 9 & 0 & 0 & 0 & 0 & 0 & 2 & 1 & 1 & 3 & 0 & 0 & 14 & 1 & 1 & 16 \\
V$_{52}$: Account Enumeration Risks. & 1 & 0 & 0 & 0 & 0 & 0 & 1 & 0 & 0 & 4 & 0 & 0 & 6 & 0 & 0 & 6 \\
V$_{53}$: Use of Compromised Credentials. & 1 & 0 & 0 & 0 & 0 & 0 & 2 & 0 & 0 & 3 & 0 & 0 & 6 & 0 & 0 & 6 \\
V$_{54}$: Misconfigured Identity Federation. & 6 & 0 & 0 & 0 & 0 & 0 & 1 & 0 & 0 & 11 & 0 & 0 & 18 & 0 & 0 & 18 \\
V$_{55}$: Excessive Privilege Provisions. & 30 & 2 & 1 & 0 & 0 & 0 & 10 & 7 & 0 & 0 & 0 & 0 & 40 & 9 & 1 & 50 \\
V$_{56}$: Insufficient Token Invalidation. & 0 & 0 & 0 & 1 & 0 & 0 & 0 & 1 & 0 & 10 & 1 & 0 & 11 & 2 & 0 & 13 \\
V$_{57}$: Insecure Token Storage. & 10 & 0 & 0 & 0 & 0 & 0 & 2 & 1 & 0 & 1 & 0 & 0 & 13 & 1 & 0 & 14 \\
V$_{58}$: Hardcoded Credentials. & 7 & 0 & 0 & 0 & 0 & 0 & 2 & 1 & 1 & 1 & 0 & 0 & 10 & 1 & 1 & 12 \\
V$_{59}$: Password Reuse. & 6 & 1 & 1 & 0 & 0 & 0 & 3 & 3 & 0 & 1 & 1 & 0 & 10 & 5 & 1 & 16 \\
V$_{60}$: Insecure Password Recovery. & 1 & 0 & 0 & 0 & 0 & 0 & 1 & 1 & 0 & 3 & 0 & 0 & 5 & 1 & 0 & 6 \\
V$_{61}$: Lack of Multi-Factor Authentication. & 3 & 0 & 0 & 0 & 0 & 0 & 2 & 1 & 0 & 6 & 0 & 0 & 11 & 1 & 0 & 12 \\
V$_{62}$: Phishing Vulnerability. & 1 & 0 & 0 & 1 & 0 & 0 & 1 & 1 & 1 & 5 & 0 & 1 & 8 & 1 & 2 & 11 \\
V$_{63}$: Unenforced Access Controls. & 8 & 0 & 0 & 0 & 1 & 0 & 3 & 0 & 0 & 5 & 0 & 0 & 16 & 1 & 0 & 17 \\
V$_{64}$: Access Grant Misconfigurations. & 3 & 0 & 0 & 0 & 0 & 0 & 1 & 1 & 1 & 5 & 0 & 0 & 9 & 1 & 1 & 11 \\
V$_{65}$: Direct Object Reference Vulnerability. & 28 & 1 & 1 & 6 & 1 & 0 & 9 & 6 & 5 & 10 & 1 & 0 & 53 & 9 & 6 & 68 \\
V$_{66}$: Over-privileged APIs. & 3 & 1 & 1 & 2 & 0 & 0 & 4 & 1 & 0 & 5 & 0 & 1 & 14 & 2 & 2 & 18 \\
V$_{67}$: Session Hijacking. & 3 & 2 & 0 & 0 & 0 & 0 & 4 & 0 & 1 & 13 & 0 & 1 & 20 & 2 & 2 & 24 \\
V$_{68}$: Cross-Site Request Forgery. & 7 & 2 & 0 & 1 & 0 & 0 & 2 & 3 & 0 & 7 & 0 & 1 & 17 & 5 & 1 & 23 \\
V$_{69}$: Session Control Exploitation. & 4 & 2 & 0 & 1 & 0 & 0 & 4 & 4 & 1 & 4 & 1 & 0 & 13 & 7 & 1 & 21 \\
V$_{70}$: Improper Session Expiry. & 6 & 1 & 1 & 0 & 0 & 0 & 2 & 3 & 1 & 7 & 0 & 0 & 15 & 4 & 2 & 21 \\

V$_{71}$: Container Misconfigurations. & 0 & 0 & 0 & 1 & 0 & 0 & 0 & 0 & 1 & 6 & 0 & 0 & 7 & 0 & 1 & 8 \\
V$_{72}$: Improper Container Isolation. & 0 & 0 & 0 & 2 & 0 & 0 & 0 & 0 & 1 & 6 & 0 & 0 & 8 & 0 & 1 & 9 \\
V$_{73}$: Direct Storage of Sensitive Data on Container Image. & 4 & 0 & 0 & 0 & 0 & 0 & 1 & 0 & 1 & 9 & 0 & 0 & 14 & 0 & 1 & 15 \\
V$_{74}$: Outdated/Insecure Container Image Usage. & 0 & 0 & 0 & 0 & 1 & 0 & 0 & 0 & 3 & 3 & 1 & 0 & 3 & 2 & 3 & 8 \\
V$_{75}$: Misconfigured Orchestration Dashboards. & 0 & 0 & 0 & 1 & 0 & 0 & 0 & 0 & 1 & 4 & 0 & 2 & 5 & 0 & 3 & 8 \\
V$_{76}$: Unrestricted Orchestration API Access. & 1 & 0 & 0 & 0 & 0 & 0 & 0 & 0 & 0 & 0 & 0 & 0 & 1 & 0 & 0 & 1 \\
V$_{77}$: Improper permissions. & 11 & 0 & 0 & 0 & 0 & 0 & 4 & 1 & 0 & 0 & 0 & 0 & 15 & 1 & 0 & 16 \\
V$_{78}$: Inherent Orchestration Tool Vulnerabilities. & 4 & 0 & 0 & 2 & 0 & 0 & 1 & 0 & 1 & 7 & 0 & 0 & 14 & 0 & 1 & 15 \\
V$_{79}$: Insecure Service Configuration. & 1 & 0 & 0 & 0 & 0 & 0 & 0 & 0 & 0 & 0 & 0 & 0 & 1 & 0 & 0 & 1 \\
V$_{80}$: Lack of Configuration Validation. & 3 & 0 & 0 & 0 & 0 & 0 & 0 & 0 & 0 & 0 & 0 & 0 & 3 & 0 & 0 & 3 \\
V$_{81}$: Embedded Secrets in Configuration Files. & 7 & 1 & 0 & 0 & 0 & 0 & 1 & 1 & 0 & 0 & 0 & 0 & 8 & 2 & 0 & 10 \\
V$_{82}$: Insecure Configuration File Validation. & 9 & 0 & 0 & 1 & 0 & 0 & 1 & 4 & 2 & 6 & 1 & 0 & 17 & 5 & 2 & 24 \\
V$_{83}$: Integration of Vulnerable Components. & 3 & 0 & 0 & 0 & 0 & 0 & 2 & 3 & 0 & 4 & 0 & 1 & 9 & 3 & 1 & 13 \\
V$_{84}$: Use of Outdated Dependencies. & 2 & 0 & 0 & 1 & 0 & 0 & 0 & 1 & 2 & 3 & 1 & 0 & 6 & 2 & 2 & 10 \\
V$_{85}$: Insufficient Dependency Scanning. & 1 & 0 & 0 & 0 & 0 & 0 & 0 & 0 & 0 & 0 & 0 & 0 & 1 & 0 & 0 & 1 \\
V$_{86}$: Lack of Transitive Dependency Validation. & 7 & 0 & 0 & 0 & 0 & 0 & 3 & 4 & 0 & 0 & 0 & 0 & 10 & 4 & 0 & 14 \\
V$_{87}$: Inconsistent Security Practices. & 0 & 0 & 0 & 0 & 0 & 0 & 0 & 0 & 1 & 0 & 0 & 0 & 0 & 0 & 1 & 1 \\
V$_{88}$: Issues in Specific Libraries. & 0 & 0 & 0 & 0 & 0 & 0 & 0 & 0 & 1 & 0 & 0 & 0 & 0 & 0 & 1 & 1 \\
V$_{89}$: Platform Misconfigurations. & 0 & 0 & 0 & 0 & 0 & 0 & 0 & 0 & 1 & 0 & 0 & 0 & 0 & 0 & 1 & 1 \\
V$_{90}$: Patch Management Complexity. & 0 & 0 & 0 & 0 & 0 & 0 & 0 & 0 & 1 & 0 & 0 & 0 & 0 & 0 & 1 & 1 \\
V$_{91}$: Legacy System Integration Vulnerabilities. & 0 & 0 & 0 & 0 & 0 & 0 & 0 & 0 & 1 & 0 & 0 & 0 & 0 & 0 & 1 & 1 \\
V$_{92}$: Mismatched Data Formats. & 0 & 0 & 0 & 0 & 0 & 0 & 0 & 1 & 0 & 0 & 0 & 0 & 0 & 1 & 0 & 1 \\
V$_{93}$: Service Mesh Configuration Errors. & 0 & 0 & 0 & 0 & 0 & 0 & 0 & 1 & 0 & 0 & 0 & 0 & 0 & 1 & 0 & 1 \\
V$_{94}$: Inconsistent Integration Security. & 4 & 1 & 1 & 0 & 0 & 0 & 4 & 2 & 0 & 0 & 0 & 0 & 8 & 3 & 1 & 12 \\
V$_{95}$: Compromised Supply Chains. & 0 & 0 & 0 & 0 & 0 & 0 & 0 & 0 & 1 & 0 & 0 & 0 & 0 & 0 & 1 & 1 \\
V$_{96}$: Third-Party Service Outages. & 6 & 0 & 0 & 0 & 0 & 0 & 3 & 2 & 0 & 0 & 0 & 0 & 9 & 2 & 0 & 11 \\
V$_{97}$: Insecure Third-Party Components. & 6 & 0 & 0 & 0 & 0 & 0 & 0 & 3 & 0 & 0 & 0 & 0 & 6 & 3 & 0 & 9 \\
V$_{98}$: Poor Security Practices in Third-Party Components. & 0 & 0 & 0 & 0 & 0 & 0 & 0 & 0 & 1 & 0 & 0 & 0 & 0 & 0 & 1 & 1 \\
V$_{99}$: Injection Vulnerabilities. & 48 & 4 & 6 & 8 & 1 & 1 & 16 & 12 & 15 & 32 & 2 & 3 & 104 & 19 & 25 & 148 \\
V$_{100}$: Improper XSS Preventions. & 0 & 0 & 0 & 12 & 3 & 2 & 0 & 0 & 14 & 36 & 1 & 0 & 48 & 4 & 16 & 68 \\
V$_{101}$: Insecure Deserialization. & 0 & 0 & 0 & 0 & 0 & 0 & 0 & 0 & 0 & 1 & 0 & 0 & 1 & 0 & 0 & 1 \\
V$_{102}$: Frameworks-Specific Flaws. & 0 & 0 & 0 & 0 & 0 & 0 & 0 & 0 & 2 & 0 & 0 & 0 & 0 & 0 & 2 & 2 \\
V$_{103}$: Cloud Misconfiguration. & 0 & 0 & 0 & 2 & 0 & 1 & 0 & 0 & 1 & 1 & 0 & 0 & 3 & 0 & 2 & 5 \\
V$_{104}$: Misconfigured Infrastructure Tools. & 0 & 0 & 0 & 1 & 0 & 0 & 0 & 0 & 1 & 0 & 0 & 0 & 1 & 0 & 1 & 2 \\
V$_{105}$: VMs Mismanagement. & 0 & 0 & 0 & 1 & 0 & 0 & 0 & 0 & 1 & 2 & 0 & 0 & 3 & 0 & 1 & 4 \\
V$_{106}$: Vulnerable Heterogeneous Networks. & 0 & 0 & 0 & 0 & 0 & 0 & 0 & 0 & 2 & 0 & 0 & 0 & 0 & 0 & 2 & 2 \\
V$_{107}$: Vulnerable CI/CD Tools. & 0 & 0 & 0 & 0 & 0 & 0 & 0 & 0 & 1 & 0 & 0 & 0 & 0 & 0 & 1 & 1 \\
V$_{108}$: CI/CD Misconfigurations. & 0 & 0 & 0 & 1 & 0 & 0 & 0 & 0 & 1 & 1 & 0 & 0 & 2 & 0 & 1 & 3 \\
V$_{109}$: Lack of Pipeline Security Controls. & 0 & 0 & 0 & 0 & 0 & 0 & 0 & 0 & 1 & 0 & 0 & 0 & 0 & 0 & 1 & 1 \\
V$_{110}$: Deploying Vulnerable Code. & 0 & 0 & 0 & 0 & 0 & 0 & 0 & 0 & 1 & 0 & 0 & 0 & 0 & 0 & 1 & 1 \\
V$_{111}$: Insecure IaC Scripts. & 1 & 0 & 0 & 1 & 0 & 0 & 0 & 0 & 1 & 2 & 0 & 0 & 4 & 0 & 1 & 5 \\
V$_{112}$: Misconfigured Automated Deployment. & 0 & 0 & 0 & 1 & 0 & 0 & 0 & 0 & 2 & 3 & 0 & 1 & 4 & 0 & 3 & 7 \\
V$_{113}$: Hardcoded Secrets in IaC. & 6 & 0 & 0 & 0 & 0 & 0 & 1 & 0 & 0 & 0 & 0 & 0 & 7 & 0 & 0 & 7 \\
V$_{114}$: No IaC Version Control. & 0 & 0 & 0 & 0 & 0 & 0 & 0 & 0 & 1 & 0 & 0 & 0 & 0 & 0 & 1 & 1 \\
V$_{115}$: Poor Secrets Rotation. & 0 & 0 & 0 & 0 & 0 & 0 & 0 & 0 & 1 & 0 & 0 & 0 & 0 & 0 & 1 & 1 \\
V$_{116}$: Lack of Centralized Secrets Management. & 0 & 0 & 0 & 0 & 0 & 0 & 0 & 0 & 1 & 0 & 0 & 0 & 0 & 0 & 1 & 1 \\
V$_{117}$: Secrets in Version Control. & 5 & 0 & 0 & 0 & 0 & 0 & 1 & 0 & 0 & 0 & 0 & 0 & 6 & 0 & 0 & 6 \\
V$_{118}$: Unmonitored Secrets Access. & 0 & 0 & 0 & 0 & 0 & 0 & 0 & 0 & 2 & 0 & 0 & 0 & 0 & 0 & 2 & 2 \\
V$_{119}$: Lack of Automated Testing. & 0 & 0 & 0 & 0 & 0 & 0 & 0 & 0 & 1 & 0 & 0 & 0 & 0 & 0 & 1 & 1 \\
V$_{120}$: Poor Manual Security Reviews. & 0 & 0 & 0 & 0 & 0 & 0 & 0 & 0 & 1 & 0 & 0 & 0 & 0 & 0 & 1 & 1 \\
V$_{121}$: Ignoring Test Results. & 0 & 0 & 0 & 0 & 0 & 0 & 0 & 0 & 1 & 0 & 0 & 0 & 0 & 0 & 1 & 1 \\
V$_{122}$: Outdated Testing Tools. & 0 & 0 & 0 & 0 & 0 & 0 & 0 & 0 & 1 & 0 & 0 & 0 & 0 & 0 & 1 & 1 \\
V$_{123}$: Inadequate Security Training. & 0 & 0 & 0 & 0 & 0 & 1 & 0 & 0 & 0 & 1 & 0 & 0 & 1 & 0 & 1 & 2 \\
V$_{124}$: Neglecting Early Security. & 0 & 0 & 0 & 0 & 0 & 0 & 0 & 0 & 0 & 1 & 0 & 0 & 1 & 0 & 0 & 1 \\
V$_{125}$: Inconsistent Security Practices. & 0 & 0 & 0 & 0 & 0 & 0 & 0 & 0 & 0 & 1 & 0 & 0 & 1 & 0 & 0 & 1 \\
V$_{126}$: Poor Security-DevOps Collaboration. & 0 & 0 & 0 & 0 & 0 & 1 & 0 & 0 & 0 & 1 & 0 & 0 & 1 & 0 & 1 & 2 \\

\hline
 \multirow{2}{*}{\textbf{\scriptsize Total Vulnerability Counts}} & 545 & 49 & 24 & 96 & 17 & 15 & 208 & 127 & 142 & 390 & 21 & 33 & 1239 & 214 & 214 & \multirow{2}{*}{\textbf{\scriptsize 1667}} \\
 \cline{2-16}
 & \multicolumn{3}{c|}{\textbf{618}} & \multicolumn{3}{c|}{\textbf{128}} & \multicolumn{3}{c|}{\textbf{477}} & \multicolumn{3}{c|}{\textbf{444}} & \multicolumn{3}{c|}{\textbf{1667}} &  \\
 \hline
 \end{longtable}
}
\twocolumn


\begin{figure*}[!t]
\centering
    \includegraphics[width=0.95\textwidth]{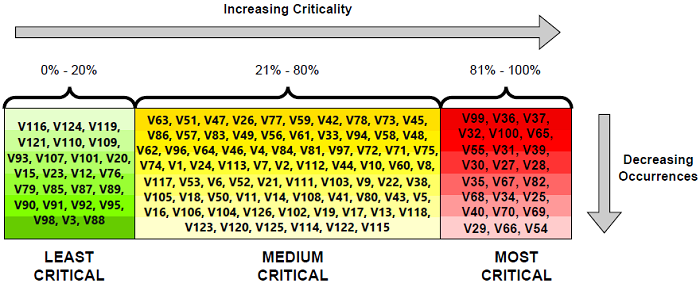}
    \caption{Vulnerability Criticality Ratings.}
    \label{fig: vcr}
\end{figure*}


\noindent ilities. Similarly, CVE-2021-21341, caused by improper access control, is mapped to V$_{40}$, which addresses excessive access privileges. This mapping process is applied to all 1,667 vulnerabilities identified. Although the mapping process is conducted with an effort to minimize bias, potential subjective interpretations could still occur. To address this, peer reviews and validation steps were implemented to ensure accuracy and consistency. By following this process, we validate the vulnerabilities defined within the proposed taxonomy.

\subsection{Mapping Analysis}

Table \ref{tab: mapping} illustrates the alignment of proposed 126 taxonomy vulnerabilities with empirically identified vulnerabilities across four benchmark applications and three scanning tools (i.e., SNYK, Trivy, and OWASP Dependency Check). Our analysis of these findings is divided into three areas: vulnerability-level analysis, application-level analysis, and tool-level analysis.

\textbf{Vulnerability-Level Analysis.} Table \ref{tab: mapping} categorizes the proposed taxonomy vulnerabilities based on their associated empirically identified vulnerabilities' frequency of occurrence across the evaluated benchmark applications. This classification provides a direct validation of the proposed vulnerabilities by highlighting their prevalence. The final column of Table \ref{tab: mapping} presents the numerical distribution of each proposed taxonomy vulnerability across all four benchmark applications and three scanning tools. For instance, V$_{99}$ appears 148 times, indicating it is the most frequently occurring vulnerability within the proposed taxonomy. Conversely, vulnerabilities such as V$_{3}$ and V$_{76}$ occur only once across the board, suggesting that some vulnerabilities are significantly more common than others. This distribution underscores the varying impact of each proposed vulnerability in real-world microservice applications.

Through this vulnerability-level analysis, we establish a criticality rating for the proposed taxonomy vulnerabilities. This rating, derived from the frequency of occurrences recorded in the mapping table, highlights the relative importance of addressing each vulnerability. Figure \ref{fig: vcr} represents this criticality rating by categorizing vulnerabilities into three distinct levels: Least Critical, Medium Critical, and Most Critical. This categorization aligns with standard practices where the top 20\% of vulnerabilities by occurrence are considered the most critical, while the bottom 20\% are deemed least critical. This approach helps prioritize which vulnerabilities require the most urgent attention. In the Most Critical category, V$_{99}$ stands out with the highest number of occurrences, totaling 148, while V$_{54}$ has the lowest number of occurrences in this category, at 18. This distribution is clearly illustrated in the figure, with varying shades of red indicating the decreasing frequency of these vulnerabilities. Similarly, within the Medium Critical category, V$_{63}$ is the most frequently occurring vulnerability, with 17 instances, with occurrences decreasing further down the list. This trend is represented in the figure by progressively lighter shades of yellow. For the Least Critical category, the color intensity increases as the number of occurrences decreases. This pattern reflects that as the frequency of a vulnerability's occurrence diminishes, it signifies improved security against these vulnerabilities. This visual representation underscores the varying levels of criticality among the proposed taxonomy vulnerabilities, aiding in the prioritization of mitigation efforts. This comprehensive criticality rating and its graphical representation provide a clear and actionable framework for addressing vulnerabilities in microservice architectures, emphasizing the need to focus on those with higher occurrences while recognizing the relative security improvements associated with lower-frequency vulnerabilities.

\textbf{Application-Level Analysis.} We conduct a detailed examination of the proposed taxonomy vulnerabilities by analyzing their association with empirically identified vulnerabilities across each benchmark microservice application. As previously detailed, a total of 1667 vulnerabilities were identified through empirical analysis, whose distribution among the four benchmark applications is as follows: 618 vulnerabilities in the Book Management System, 128 in the Cloud-Native E-Shop Application, 477 in the Spring Boot Application, and 444 in the Knowledge Base Application. 

Table \ref{tab: mapping} provides a comprehensive view of these vulnerabilities, allowing us to assess their distribution within each application. For instance, V$_{36}$ appears most frequently within the Book Management System, with a total of 64 occurrences. Conversely, vulnerabilities such as V$_{72}$ and V$_{74}$ are not detected within this application at all, indicating a zero occurrence rate. This analysis enables us to evaluate the criticality of each proposed taxonomy vulnerability based on its frequency of occurrence within each application. A higher number of occurrences suggests a greater level of criticality. By analyzing these patterns, we can gauge the relevance and impact of the defined taxonomy across various microservice applications. This application-level analysis not only highlights the varying degrees of vulnerability impact across different applications but also reaffirms the validity and applicability of the proposed taxonomy. It demonstrates that the taxonomy is relevant and useful in identifying and addressing vulnerabilities across diverse microservice environments.

\textbf{Tool-Level Analysis.} The tool-level analysis focuses on the effectiveness and coverage of each vulnerability scanning tool, as illustrated in Table \ref{tab: mapping}. This table details how each of the three scanning tools, SNYK, Trivy, and OWASP Dependency Check, has identified vulnerabilities across the four benchmark applications. For example, SNYK has identified a total of 1,239 unique vulnerabilities across all four applications, making it the most comprehensive tool in terms of coverage. In contrast, both Trivy and OWASP Dependency Check have each identified 214 vulnerabilities. This discrepancy highlights the differing strengths and scopes of each tool. A more granular analysis reveals how specific vulnerabilities from the proposed taxonomy are detected by each tool. For instance, SNYK has detected 104 vulnerabilities related to V$_{99}$ across the four applications, demonstrating its robust detection capabilities for this category. Trivy, on the other hand, has identified 20 vulnerabilities related to V$_{36}$, while OWASP Dependency Check has detected 16 vulnerabilities pertaining to V$_{100}$. This tool-specific analysis provides valuable insights into the performance and coverage of each scanning tool. By understanding which tools are more effective at identifying certain types of vulnerabilities, we gain a clearer picture of their utility in securing microservice applications. This information is crucial for selecting appropriate tools for different security needs and improving the overall security posture of microservice architectures.

\subsection{Implications for Practitioners and Researchers}

The transition towards microservice architecture is becoming increasingly prevalent in the industry, and the insights gained from this research, particularly vulnerability mapping, offer significant implications for both practitioners and researchers. The findings and analysis presented in this study can guide future efforts to enhance security practices and tools in microservice environments.

For practitioners, this research provides valuable information on the criticality of vulnerabilities within microservice architectures. Practitioners can leverage this data to focus their attention on the most critical vulnerabilities when designing and implementing microservices. By prioritizing these high-impact vulnerabilities, practitioners can improve the security posture of their systems and applications, ensuring that the most pressing issues are addressed early in the development process. This targeted approach will help in designing more resilient systems that are better protected against prevalent and severe vulnerabilities.

For researchers, the findings from our study and the proposed vulnerability taxonomy open several promising avenues for further investigation and advancement in the field of cybersecurity for microservices. One potential avenue is to extend the evaluation of the proposed microservice vulnerability taxonomy to other benchmark applications and real-world microservice architectures. This could involve applying the taxonomy to a broader range of systems and utilizing additional vulnerability scanning tools to assess its robustness and adaptability across different microservice architectures. By doing so, researchers can validate and refine the taxonomy's applicability and effectiveness in diverse contexts, enhancing its utility as a comprehensive tool for identifying and addressing vulnerabilities. Another crucial area for future research is improving security features for microservices. The vulnerability criticality ratings derived from our study provide valuable insights into the most common and severe vulnerabilities within microservices. Researchers can leverage this information to explore strategies for mitigating these issues and developing advanced security features. Another promising future research direction is the development of an automated vulnerability management framework for microservice architectures, advancing from DevOps to DevSecOps. This framework should tackle the distinct challenges associated with the distributed and modular nature of microservices, including the need for continuous monitoring and dynamic updates. Emphasizing automation in vulnerability detection, assessment, and prioritization could enhance the overall security management and resilience of microservices.

\section{Threats to Validity}\label{sec7}

The validity of our research findings is subject to several potential threats. Firstly, our empirical evaluation is based on a limited sample size of four benchmark microservice applications. This constraint may impact the generalizability of our results, as insights derived from a small sample may not be applicable to a broader array of applications. To address this issue, we selected four diverse applications that utilize different technologies, aiming to provide a representative sample. However, the inherent limitation of a small sample size remains a consideration. Secondly, the study is confined to three specific vulnerability scanning tools — SNYK, Trivy, and OWASP Dependency Check. This selection could affect the generalizability of our findings to other tools not included in the study. Although these tools were chosen for their prominence and diverse capabilities within the cybersecurity community, the results may be particularly relevant to the tools used in this study and may not fully reflect the performance of all available scanning tools. Another threat stems from the static nature of the proposed vulnerability taxonomy. The taxonomy was developed based on the current understanding of microservice architectures and may not account for emerging vulnerabilities as technology evolves. As the landscape of microservices continues to advance, the taxonomy may become outdated. Therefore, it is crucial to view this taxonomy as a foundational framework that requires regular updates to incorporate new vulnerabilities and adapt to technological changes. Lastly, the methodology used for assigning criticality ratings to vulnerabilities — dividing them into the top 20\%, bottom 20\%, and the remaining 60\% — may not be universally applicable. While this approach serves as a solid foundation for developing a criticality scale, it may not align with specific industry requirements or the needs of individual organizations. Different contexts may necessitate alternative methods for categorizing and assessing vulnerability criticality. 

\section{Conclusion}\label{sec8}

Microservice architectures are transforming both small enterprises and large organizations by offering significant improvements in maintainability, scalability, and flexibility. However, these benefits come with their own set of security challenges. The inherent nature of microservices, characterized by containerization, inter-service communication, and orchestration, introduces unique vulnerabilities that can complicate security management. This study addresses these challenges by developing a comprehensive taxonomy of security vulnerabilities associated with microservice architectures. Through a review and analysis of 62 studies from existing literature, we identified and categorized 126 distinct vulnerabilities. To ensure the accuracy and practical applicability of our taxonomy, we conduct an empirical validation using four benchmark microservice applications and three different vulnerability scanning tools. This empirical analysis is crucial in demonstrating the real-world relevance of the identified vulnerabilities and provides a practical context for our findings. By mapping the vulnerabilities detected in our empirical analysis to our proposed taxonomy, we reveal critical insights that can guide both practitioners and researchers. This mapping not only highlights the most pressing security concerns but also provides actionable recommendations for improving the security posture of microservice architectures. For instance, researchers can explore the development of an automated vulnerability management framework tailored to microservice architectures, facilitating a transition from DevOps to DevSecOps practices. We hope the insights of this study will advance the development of more robust security measures and contribute to a deeper theoretical and practical understanding of microservice vulnerabilities.

\section*{Acknowledgment}

\begin{itemize} [leftmargin=*]

    \item The authors acknowledge the SLB management for their support, the SLB review committee for their valuable feedback, and for granting permission to publish this paper.

    \item The authors also extend their gratitude to Migrova for their collaboration, which provided insights into real-world cybersecurity vulnerabilities in microservice architectures.

\end{itemize}

\bibliographystyle{IEEEtran}
\bibliography{bibliography}

\end{document}